\documentclass[authoryear,5p,twocolumn]{elsarticle} 

\usepackage{amssymb}
\usepackage{amssymb,amsmath}
\usepackage{multirow}
\usepackage{subfigure}
\usepackage[hyphens]{url}
\usepackage{breakurl}

\usepackage{lineno}

\journal{Astronomy and Computing}

\begin{document}

\begin{frontmatter}



\title{Machine learning-based seeing estimation and prediction using multi-layer meteorological data at Dome A, Antarctica}


\author[address1,address2,address3]{Xu Hou}

\author[address4]{Yi Hu}
\ead{huyi.naoc@gmail.com}

\author[address1,address2]{Fujia Du \corref{cor1}}
\ead{fjdu@niaot.ac.cn}
\cortext[cor1]{Corresponding author}

\author[address5]{Michael C. B. Ashley}

\author[address1,address2]{Chong Pei}

\author[address4]{Zhaohui Shang}

\author[address6]{Bin Ma}

\author[address1,address2,address3]{Erpeng Wang}

\author[address1,address2,address3]{Kang Huang}

\address[address1]{National Astronomical Observatories/Nanjing Institute of Astronomical Optics \& Technology, Chinese Academy of Science, Nanjing, 210042, China}
\address[address2]{CAS Key Laboratory of Astronomical Optics \& Technology, Nanjing Institute of Astronomical Optics \& Technology, Nanjing, 210042, China}
\address[address3]{University of Chinese Academy of Sciences, Beijing, 100049, China}
\address[address4]{National Astronomical Observatories, Chinese Academy of Sciences, Beijing, 100101, China}
\address[address5]{School of Physics, University of New South Wales, Sydney, 2052, Australia}
\address[address6]{School of Physics and Astronomy, Sun Yat-Sen University, Zhuhai, 519082, China}

\begin{abstract}
Atmospheric seeing is one of the most important parameters for evaluating and monitoring an astronomical site. Moreover, being able to predict the seeing in advance can guide observing decisions and significantly improve the efficiency of telescopes. However, it is not always easy to obtain long-term and continuous seeing measurements from a standard instrument such as differential image motion monitor (DIMM), especially for those unattended observatories with challenging environments such as Dome A, Antarctica. In this paper, we present a novel machine learning-based framework for estimating and predicting seeing at a height of 8 m at Dome A, Antarctica, using only the data from a multi-layer automated weather station (AWS). In comparison with DIMM data, our estimate has a root mean square error (RMSE) of 0.18 arcsec, and the RMSE of predictions 20 minutes in the future is 0.12 arcsec for the seeing range from 0 to 2.2 arcsec. Compared with the persistence, where the forecast is the same as the last data point, our framework reduces the RMSE by 37 per cent. Our method predicts the seeing within a second of computing time, making it suitable for real-time telescope scheduling.
\end{abstract}

\begin{keyword}
 Atmospheric effects \sep Methods: data analysis \sep Methods: statistical \sep Long short-term memory  network \sep Gaussian process regression \sep  Telescopes 
\end{keyword}

\end{frontmatter}

\section{Introduction}           
\label{sect:intro}

Atmospheric seeing is one of the most critical parameters to characterize an 
optical/infrared astronomical site. The differential image
motion monitor (DIMM) is a widely used instrument for selecting a new site or monitoring the seeing at an existing observatory \citep{Sarazin90}. 
As the next-generation of extremely large telescopes come online, monitoring and predicting seeing will become critically important for scheduling these expensive facilities in real-time. Clearly, if one can, e.g.,  
predict good seeing for the next hour, the scheduler of
the telescope can assign high priorities to those programs that would benefit.

Since seeing is caused by the temporal variation of the refractive index through the line of sight, one can forecast seeing by forecasting a set of meteorological parameters. Many mature numerical models for weather forecasting have been introduced in astroclimatic investigations. For example, \cite{Masciadri13} and \cite{Lascaux13} applied a non-hydrostatic mesoscale model (MESO-NH) to reconstruct the vertical stratification of the atmospheric parameters and predict the seeing, isoplanatic angle, and wavefront coherence time for the two European Southern Observatory ground-based sites of Cerro Paranal and Cerro Armazones. For the Large Binocular Telescope, \cite{Turchi17} used the same MESO-NH model to predict the surface-layer atmospheric parameters at Mt Graham, Arizona. \cite{cherubini2008modelingb,cherubini2008modelinga} used the fifth-generation Pennsylvania State University–NCAR Mesoscale Model (MM5) to calculate the turbulent fluctuations of the atmospheric refractive index and seeing for the summit of Maunakea. Weather Research and Forecasting (WRF) is another non-hydrostatic mesoscale model, which is also widely used in forecasting seeing. \cite{cherubini2011operational} used the WRF to predict seeing at the Maunakea Observatories. \cite{Giordano13} used the WRF to forecast seeing at the Observatorio del Roque de Los Muchachos (ORM). \cite{Qian2021} used the WRF model to investigate the atmospheric optical turbulence for the Chinese Twelve Meter Telescope candidate site at the Ali observatory, Tibet. However, these numeric weather prediction (NWP) models can only forecast the meteorological parameters, such as the vertical profiles of the temperature, the wind speed, and direction, etc. To derive the seeing we need to know the refractive index structure constant $C_n^2(h)$ from the output of the models. This can be done using parametrized schemes such as those developed by \cite{Masciadri99}, \cite{Trinquet07}, and \cite{cherubini2013Another}. Therefore, the accuracy of seeing predictions is strongly dependent on both NWP models and the techniques used to derive seeing from the models. 

Dome A, located at S80$^\circ$25$^\prime$, E77$^\circ$05$^\prime$, 4089 m above 
sea-level \citep{Hu14}, is the highest place on the Antarctica plateau. 
Long before it was first visited by the 21st Chinese National Antarctica Research Expedition (CHINARE) team in 2005, 
Dome A had been considered as one of the best ground-based astronomical sites over a broad range of the electromagnetic spectrum
\citep{Marks02}. Beginning with the 24th CHINARE visit to Dome A there have been astronomical activities at the site, and many 
instruments have been deployed for site testing and science \citep{Yang09}. 
The results from these instruments have revealed the excellent site 
conditions of Dome A. 
For example, \cite{Bonner10} reported a median turbulent boundary layer thickness of 13.9 m at Dome A using sonic radar data between February and August 2009. \cite{Hu14,Hu19} found that a strong temperature inversion existed just above the ground surface and could last for tens of hours, using data from multi-layer experiments known as the Kunlun Automated Weather Station (KLAWS and KLAWS-2G). By directly measuring the seeing from a DIMM instrument called Kunlun-DIMM (KL-DIMM),  \cite{Ma20a} reported a superb median free-atmospheric seeing of 0.31 arcsec at Dome A, which could be achieved at a height of only 8 m. More details of the progress in astronomy and site testing at Dome A can be found in \citet{Shang20} and \citet{Ma20b}.

However, operating a DIMM at Dome A and obtaining long-term continuous seeing measurements is challenging because of the extremely low temperatures, the difficulty of icing of optical elements, and the need for fully unattended operation \citep{Hu14, Hu19}.
Although an indium tin oxide (ITO) layer is coated on the wedges of the two sub-apertures and used as a heater, KL-DIMM still suffers from occasional ice accumulation on \citep{Ma20a}. Consequently, we were led to explore easy-to-implement and ice-resistant equipment to estimate the seeing at Dome A. Fortunately, such 
equipment does exist. \cite{Hu14} found that the strength of the temperature inversion (i.e., the temperature
difference at two heights, driven by the radiative cooling of the ice) measured from KLAWS was strongly anti-correlated to 
the boundary layer thickness. \cite{Ma20a} found an anti-correlation between the 
strength of the temperature inversion and seeing. Additionally, the seeing was affected by the local 
wind. This makes it possible to estimate, and even forecast into the future, the seeing at Dome A 
by using data from a reliable and straightforward multi-layer AWS.  

Recently, machine learning methods have shown great promise in weather forecasting \citep{Grover15,Rasp20}. Therefore, machine learning-based methods may be an attractive approach for predicting the seeing \citep{Milli20,cherubini2022}. In this paper, we present a new dedicated method to estimate and forecast the seeing at Dome A using only data from KLAWS-2G. Instead of employing NWP models and parametrized schemes, we use a machine learning-based Long short-term memory (LSTM) network to predict the future meteorological parameters and a Gaussian process regression (GPR) method to derive the seeing from predicted meteorological parameters. We find that the root mean square error (RMSE) of our LSTM-GPR seeing estimation method is 0.18 arcsec for the whole seeing 
range for which we have observations, and our forecasts 20 minutes into the future have an RMSE of 0.12
arcsec for the seeing range from 0 to 2.2 arcsec. In Section \ref{sec:data}, we 
describe the data used in this paper and the data preprocessing procedure. In 
Section \ref{sec:model}, we present our machine learning models. In Section 
\ref{sec:res}, we compare our seeing estimations and predictions with observations. 
Finally, we present the discussion and conclusion in Section \ref{sec:Dis}.

\section{OVERVIEW OF THE DATA}
\label{sec:data}

We use the meteorological data collected from KLAWS-2G \citep{Hu19} and seeing measurements from KL-DIMM \citep{Ma20b} to construct our training and testing data. 

\begin{figure}
	\centering
	\includegraphics[width=5.5cm]{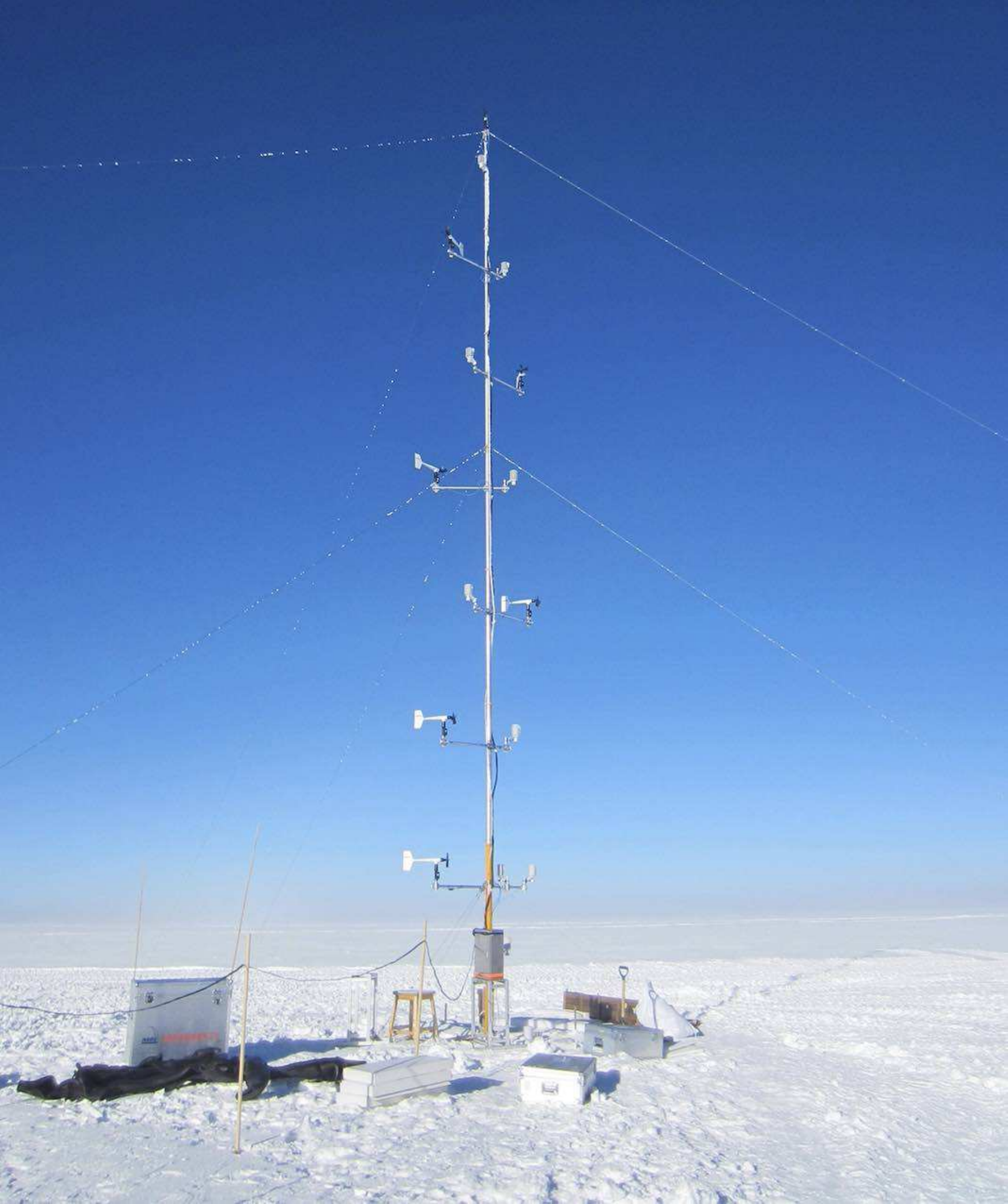}
	\caption{KLAWS-2G installed at Dome A in January 2019. 
	}
	\label{fig:KLAWS} 
\end{figure}

The first generation KLAWS was installed at Dome A in 2011 by the 27th CHINARE team, and it worked for the entire year \citep{Hu14}. In 2015, an improved version called KLAWS-2G was installed by the 31st CHINARE team. KLAWS-2G has a 15 m high mast, upon which were mounted ten temperature sensors from $-1$ m (i.e., below the ice level) to 14 m, seven wind speed and direction sensors from 2 m to 14 m, one air pressure sensor at 2 m, and one relative humidity sensor at 2 m. KLAWS-2G operated until August 2016, and then in a degraded mode with only sensors at 4 m and below from January 2017 to May 2018. In 2019, a new KLAWS-2G was installed by the 35th CHINARE team (see Fig. \ref{fig:KLAWS}). Unfortunately, it only operated from 2019 January 23 to March 15 because of a power failure. KLAWS-2G sampled all the sensors every ten seconds, and the total number of measurements was 630,000.  

\begin{figure}
	\centering
	\includegraphics[width=5.5cm]{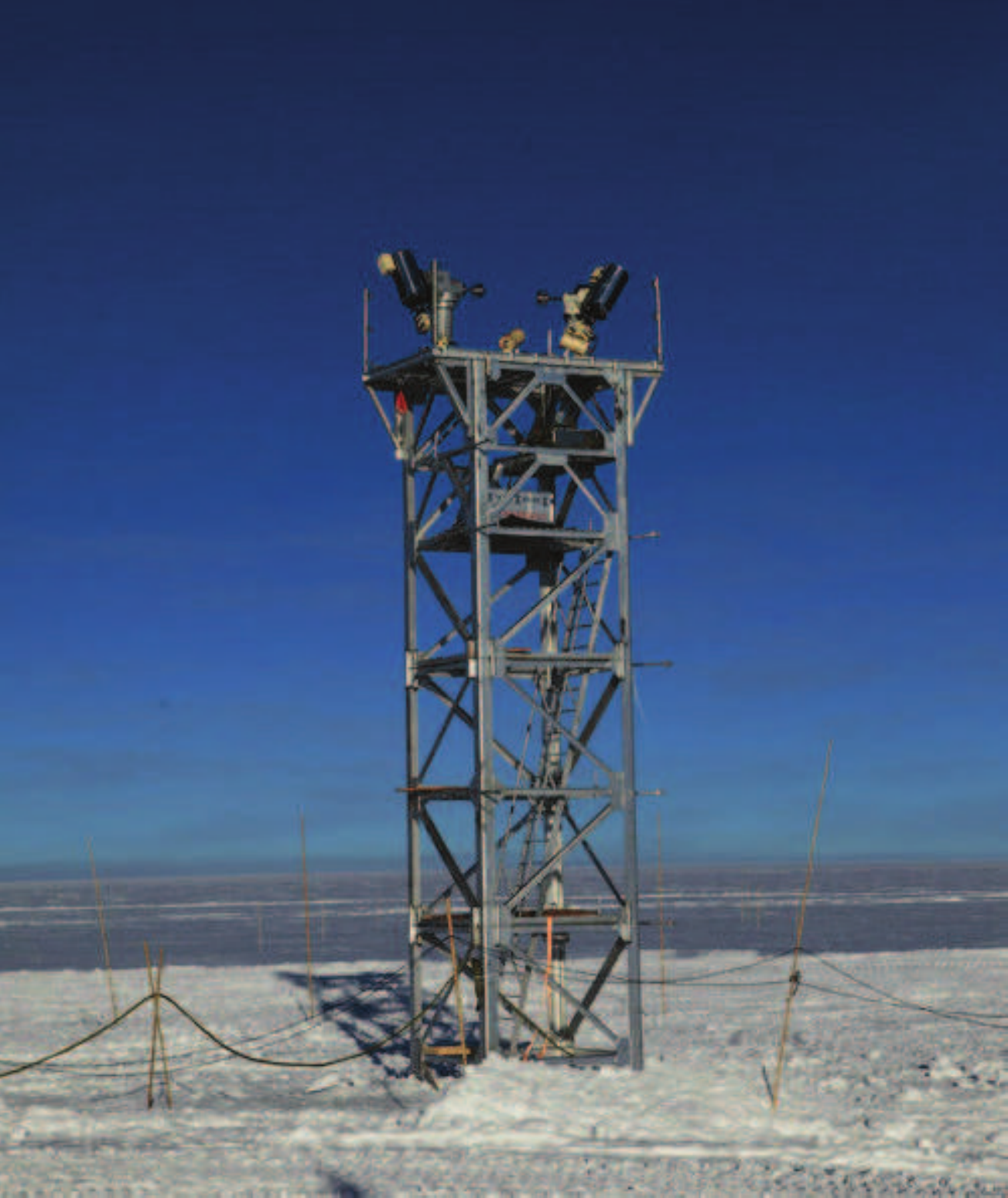}
	\caption{The two KL-DIMMs installed on a 8 m tower at Dome A in January 2019. 
	}
	\label{fig:KL-DIMM}
\end{figure}

\begin{figure}
	\centering
	\subfigure[]
	{
	 \begin{minipage}{9cm}
	  \includegraphics[width=\textwidth]{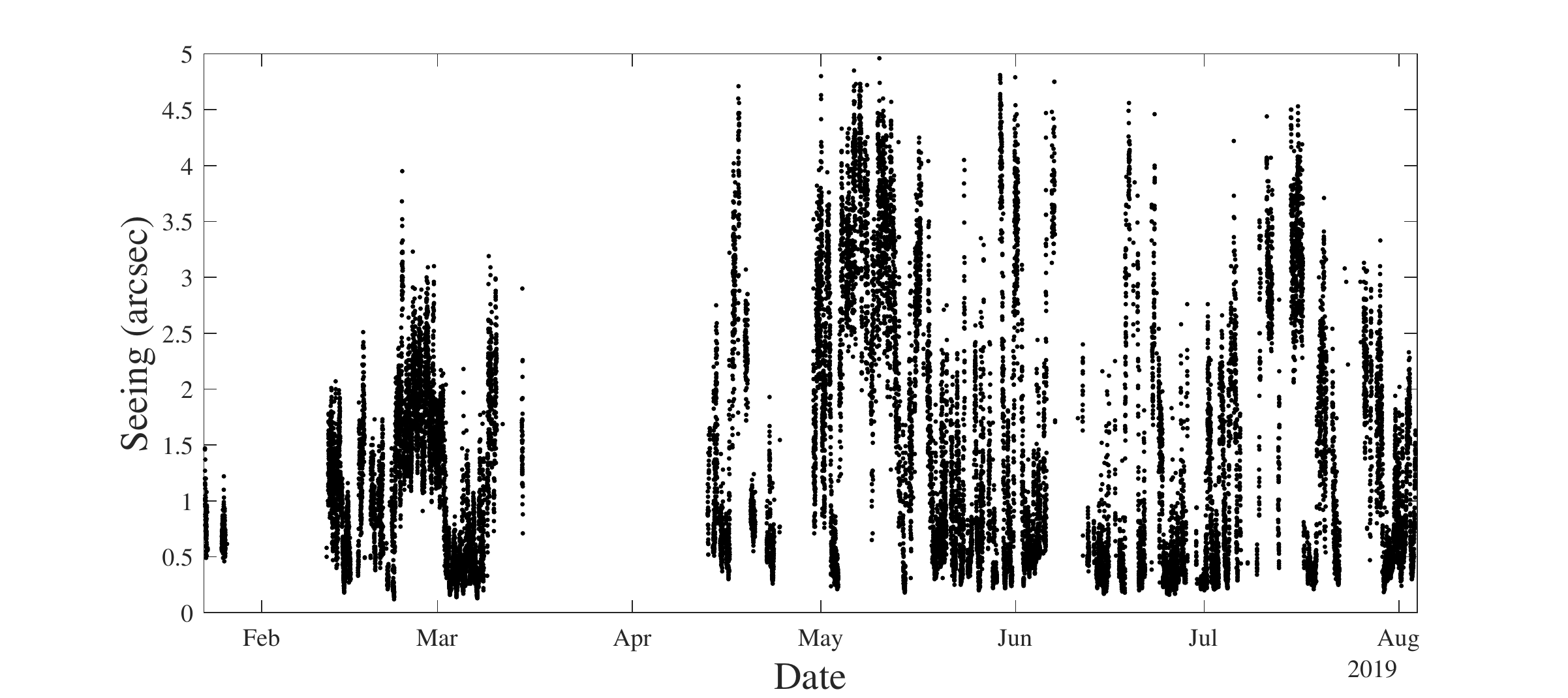}
	 \end{minipage}
	}
	\quad
	   \subfigure[]
	   {
		\begin{minipage}{9cm}
		 \includegraphics[width=\textwidth]{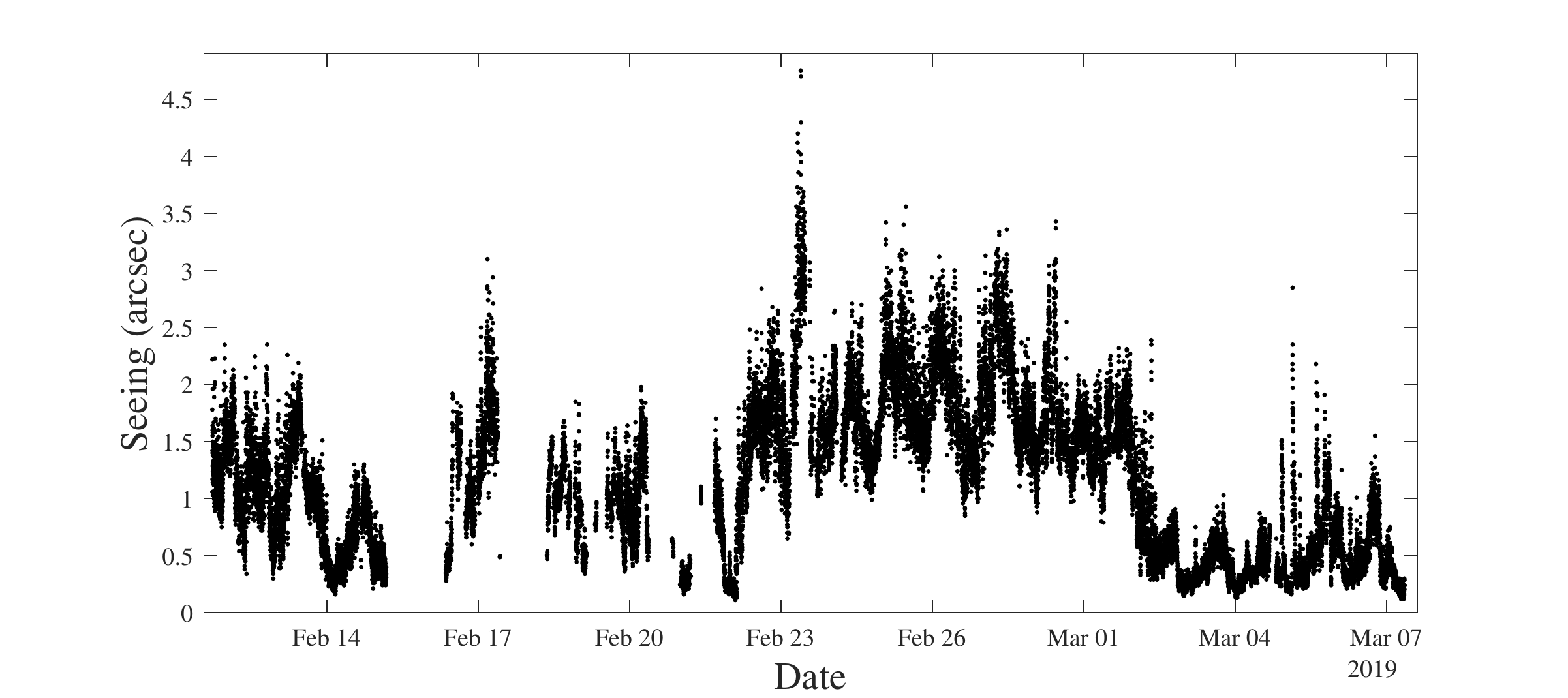}
		\end{minipage}
	   }
	   \quad
	   \subfigure[]
	   {
		\begin{minipage}{9cm}
		 \includegraphics[width=\textwidth]{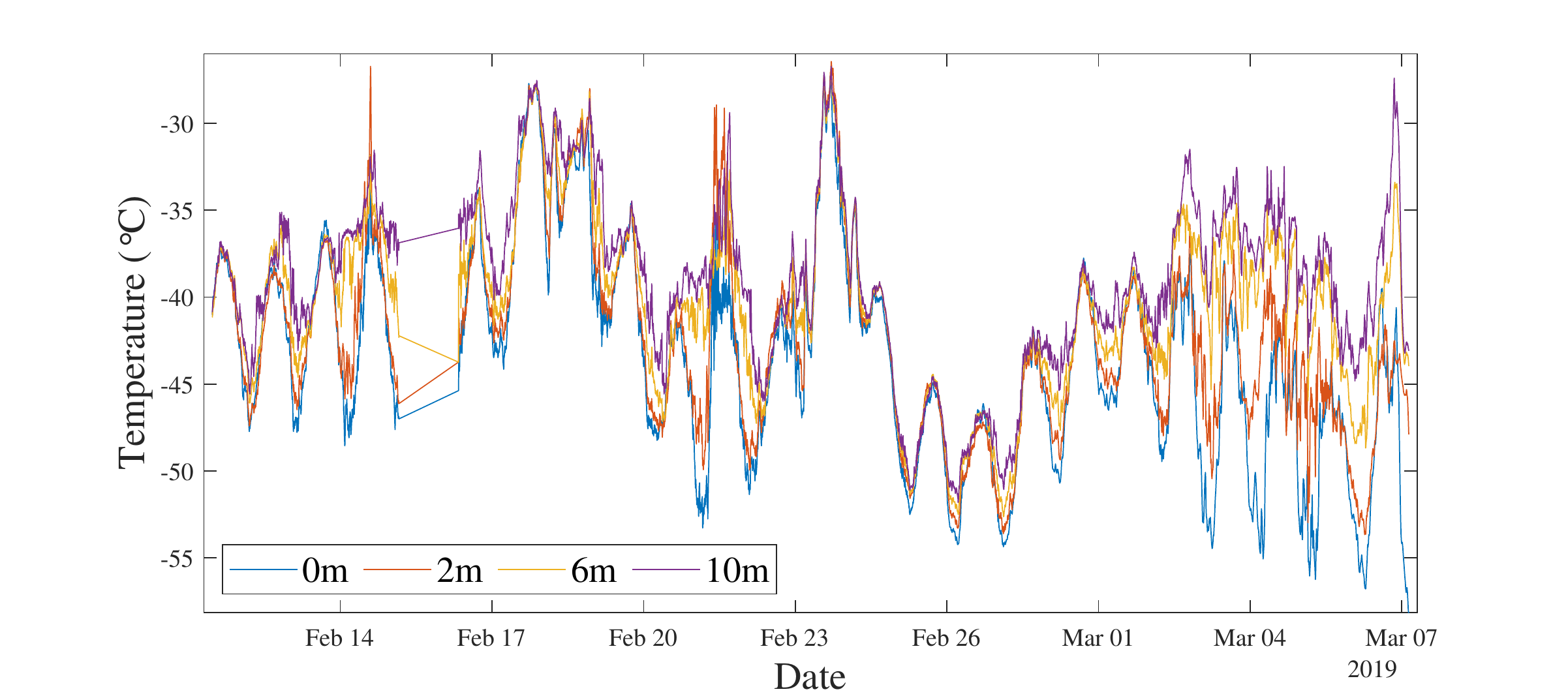}
		\end{minipage}
         }
    \quad
    \subfigure[]
	   {
		\begin{minipage}{9cm}
		 \includegraphics[width=\textwidth]{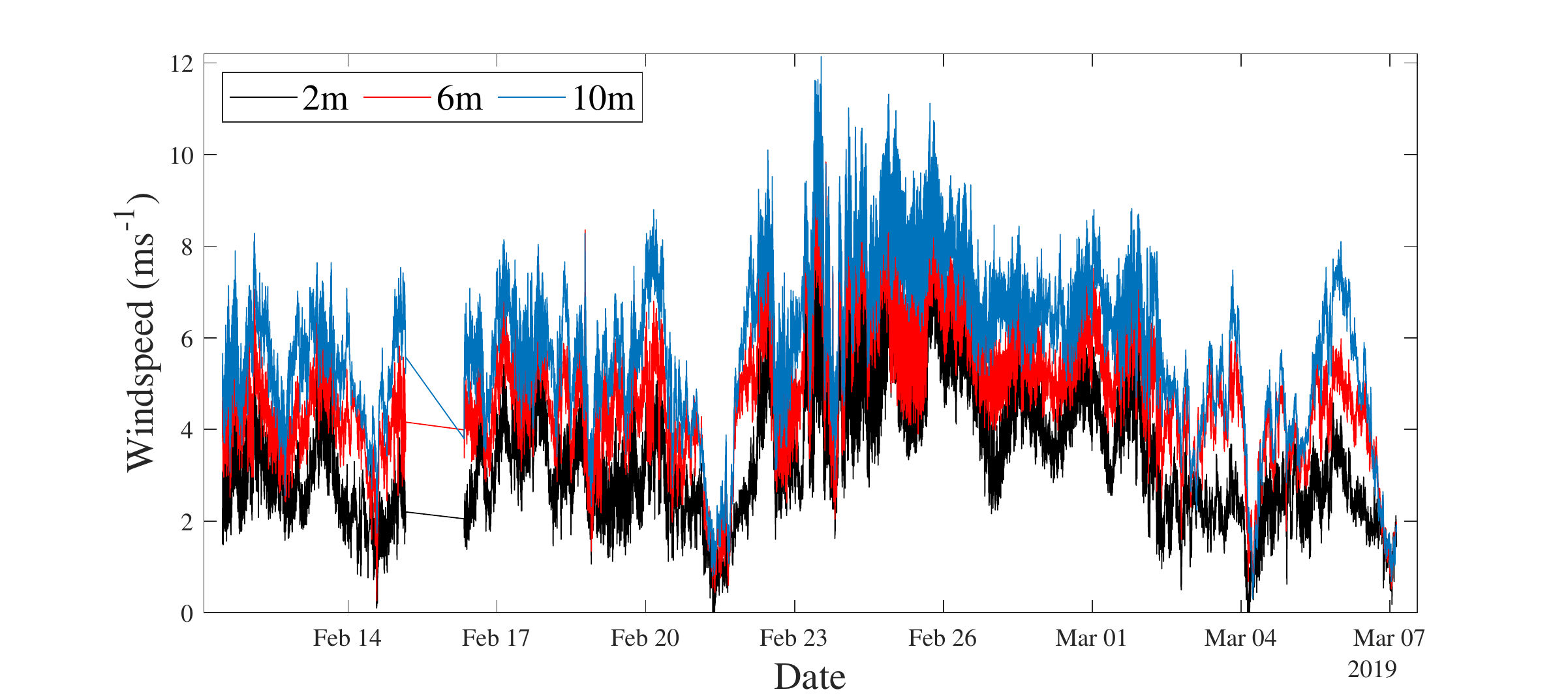}
		\end{minipage}
	   }
	   \caption{(a) Seeing data at the height of 8m in 2019. (b) Seeing data from 2019 February 11 to March 7. (c) Temperature data at heights of 0 m, 2 m, 6 m, and 10 m from 2019 February 11 to March 7. (d) Wind speed data at heights of 2 m, 6 m, and 10 m from 2019 February 11 to March 7.  
		} 
	   \label{fig: original data}
\end{figure}

Two KL-DIMMs, which were almost identical for redundancy, were installed on an 8 m tower at Dome A in 2019 (see 
Fig. \ref{fig:KL-DIMM}). 
Unfortunately, tracking and ice-accumulating problems on the left KL-DIMM resulted in little data from this instrument. Therefore, the
seeing data we used in this paper are from the righthand one. KL-DIMM generated a 
measurement every minute. It operated until 2019 August 4 and made 160,000 
measurements in 2019. However, the seeing data from March 8 to 15 were lost. 
In total, therefore, we have one month of synchronous data from KLAWS-2G and KL-DIMM for training and testing. 
The available data is shown in Fig. \ref{fig: original data}. Both KLAWS-2G and KL-DIMMs are supported by the Plateau Observatory A (PLATO-A), an automated observatory installed in 2012 \citep{Ashley10}.

\subsection{Data preprocessing}
\label{subsec:prep}  

To create clean and uniform input for our framework, we preprocessed the raw instrumental data by resampling, denoising, and rescaling. These procedures are described in detail below.  

\begin{figure}
	\centering
    	\includegraphics[width=\columnwidth]{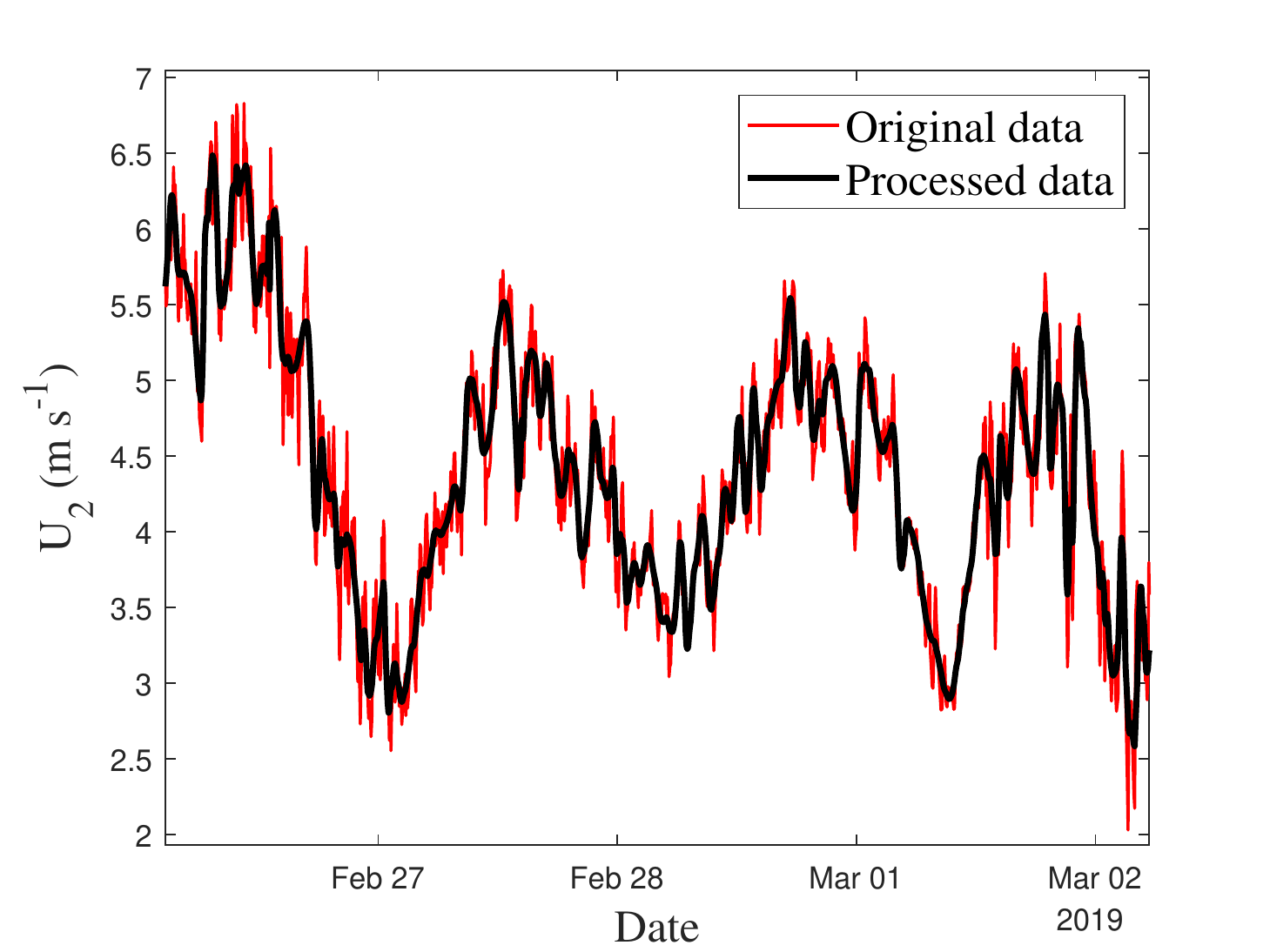}
		\caption{ Filtering high-frequency components in wind speed. 
		}
	\label{fig: data prepareation}
\end{figure}

The sampling time for the meteorological parameters was 10 seconds, while the sampling time 
for the seeing was 1 minute. However, our framework requires that all data have the 
same sampling frequency. 
Consequently, the seeing and meteorological data needed to be resampled and synchronized. 
Since the changes in some meteorological parameters are slow. The high sample rate will cause the data redundancy and the model tends to overfitting.
Therefore, the meteorological data were synchronized with the seeing data to have one data point every 5 minutes.
The detailed processes are as follows: 
\begin{itemize}
	\item [1)] 
Select samples covering a 10 minute period, and remove outliers. 
There were a few outliers in the meteorological data due to abnormal network communications and digital glitches. The data were considered outliers if the data differences exceed three times the RMSE over a 10 minute period.
	Spline interpolation was then used to replace the outliers.   
	\item [2)]
	Average the meteorological data over a precise one-minute interval based on the seeing data timestamps. Then
	compare the timestamps of the meteorological and seeing data, and sychronize to the closest timestamp.  This give a maximum synchronization error of 5 seconds. 
	\item [3)]
	Take the average of every 5 data points to obtain a new dataset with a sampling interval of 5 minutes.  
\end{itemize} 
Because of the existence of the high-frequency components in the meteorological data, a wavelet packet decomposition (WPD) method, which is optimal for analyzing time-series signals \citep{Yang16}, was applied to filter the high-frequency components. Fig. \ref{fig: data prepareation} shows the filtering results for the wind speed at 2 m.

Finally, we normalize the raw data using z-score normalization so that the data with small numerical values will not be ignored by the training network. 

\begin{figure}
	 \centering
	 \subfigure
	 {
		\begin{minipage}{8cm}
		 \includegraphics[width=\columnwidth]{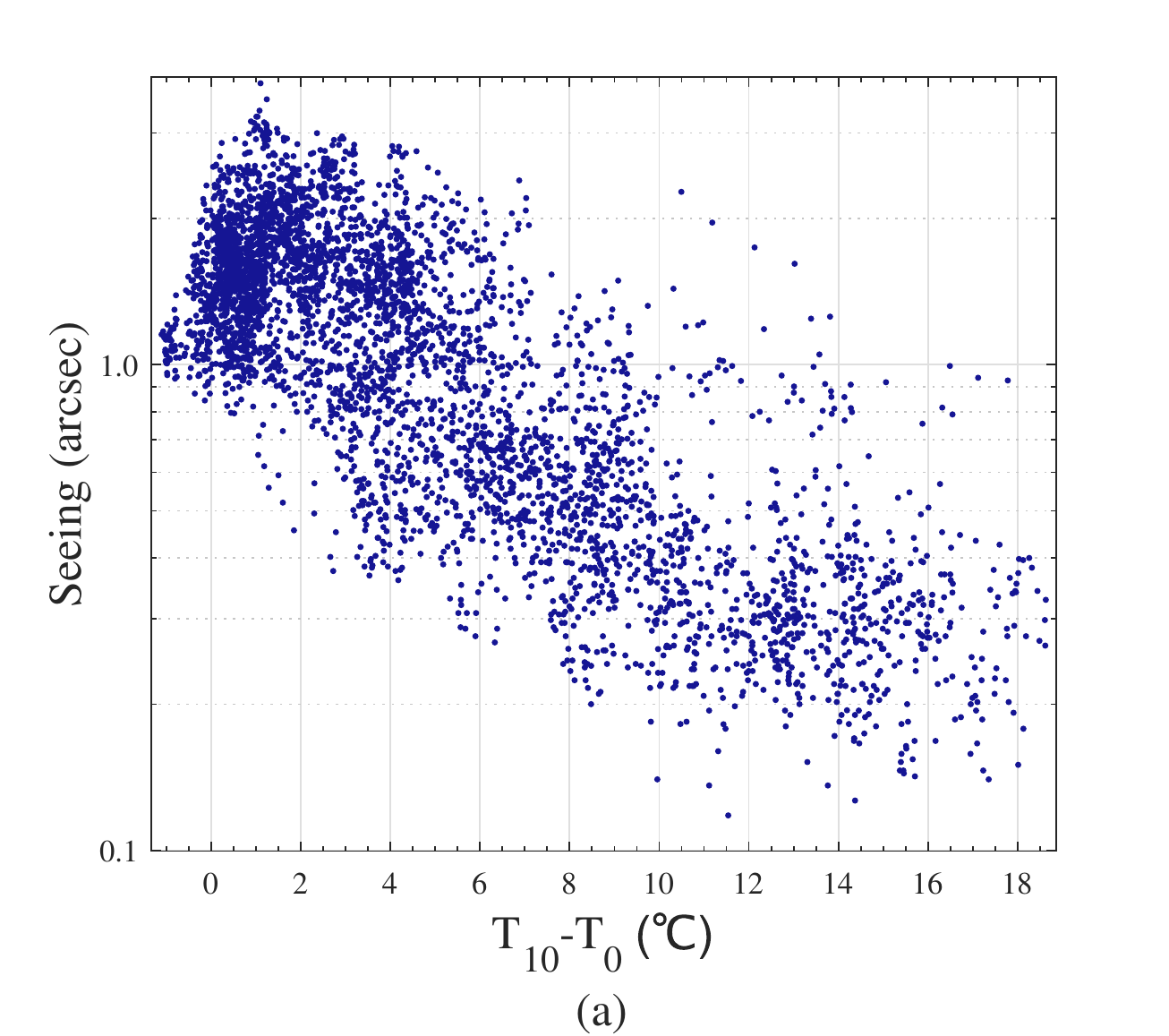}
		\end{minipage}
	  }
	 \quad
		\subfigure
		{
		 \begin{minipage}{8cm}
		\includegraphics[width=\columnwidth]{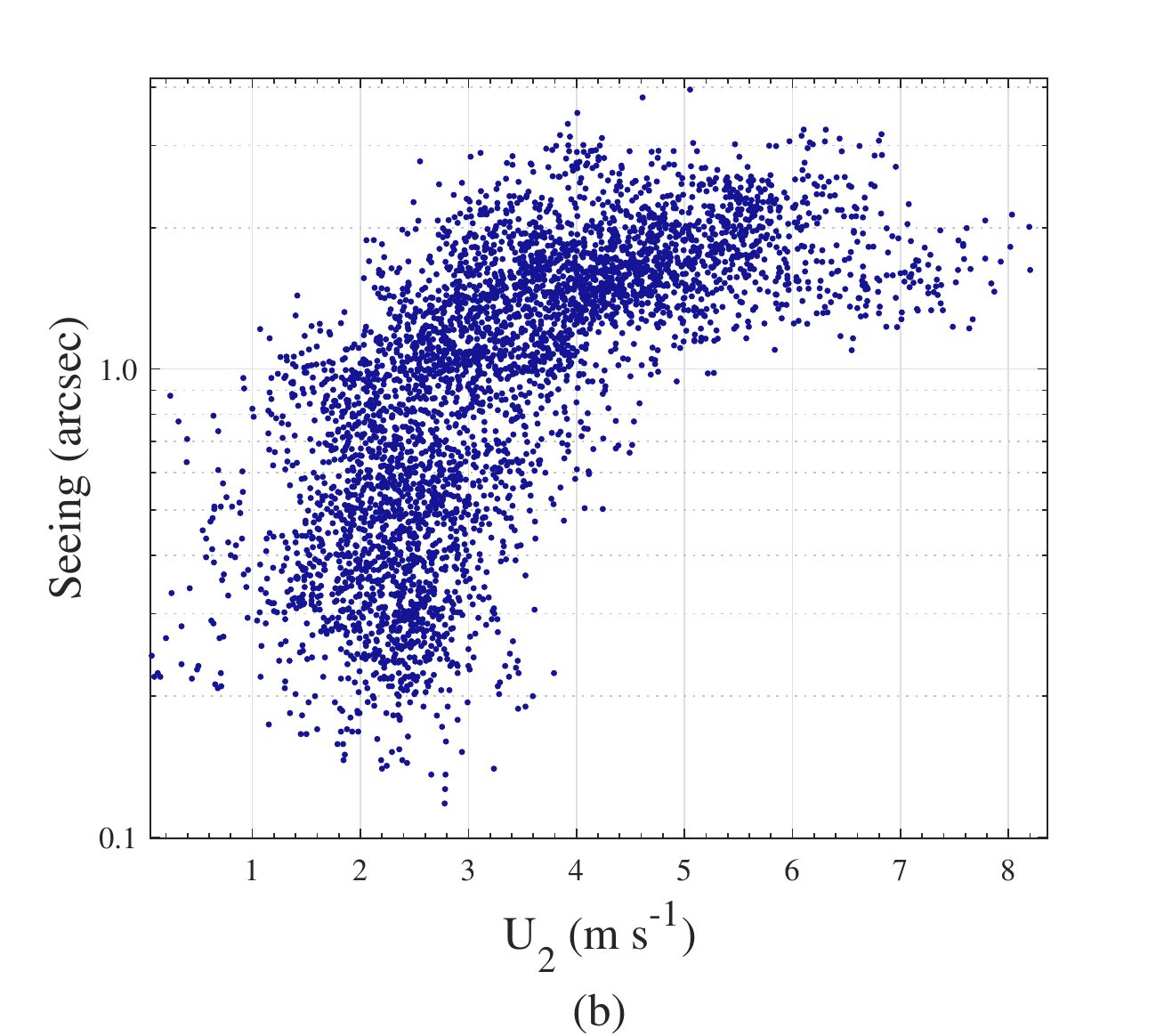}
		 \end{minipage}
		}
	\caption{The correlations between seeing and meteorological parameters. (a) The strong correlation between seeing and $T_{10}-T_{0}$, the temperature difference between sensors at heights of 10 and 0 metres. (b) $U_{2}$, the wind speed at a height of 2 m, versus seeing at 8 m.}
	\label{fig:seeing and meteorological parameter statistics}
\end{figure}

\begin{figure}
	\includegraphics[width=\columnwidth]{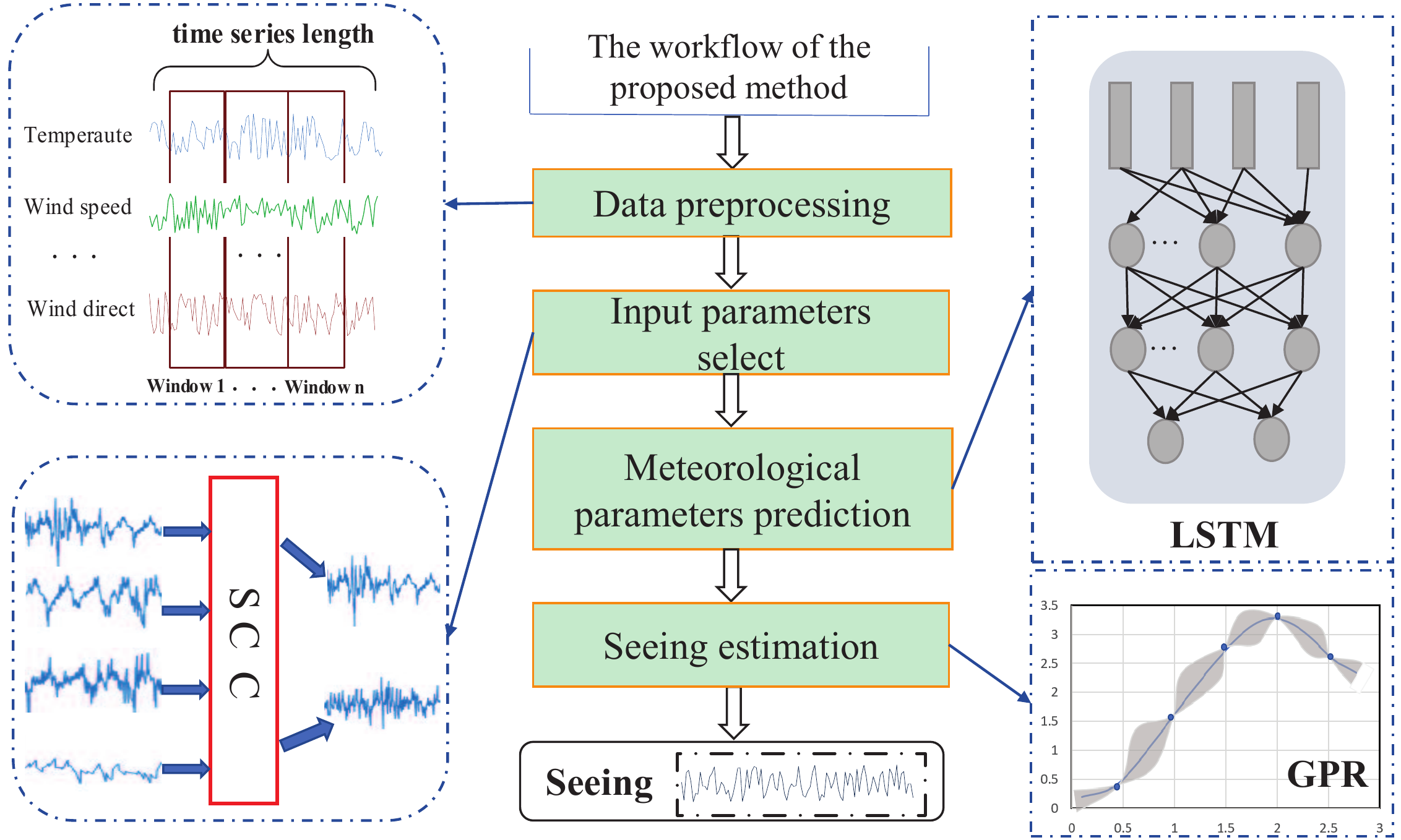}
    \caption{ Seeing estimation and prediction framework. }
	\label{fig:LSTM_predictor}
\end{figure}

\subsection{Input parameters selection}
\label{subsec:select} 

There are dozens of sensors installed on KLAWS. Some of them are designed for redundancy, such as adjacent temperature sensors and anemometers, while some of them may be irrelevant to seeing estimation and forecasting. If all the parameters are present in the input of our model, it will unnecessarily complicate the model, with little or no improvement in performance, and will increase learning and forecasting time. To select the most important parameters, we calculated the Spearman's rank correlation coefficients (SCC) \citep{press2007numerical} between all the parameters and the seeing. SCC is a nonparametric (distribution-free) rank statistic. It is a measure of a monotone association that is used to measure the strength of an association between two parameters. The mathematical formula of SCC can be expressed as follows:
\begin{gather}
	\label{SCC}
	\rho = 1 - \frac{6\sum d_{i}^2} {n(n^2 - 1)}, 
\end{gather}
where $n$ is the number of meteorological parameters and $d_{i}$ is the difference between the ranks of the seeing and meteorological parameters. The results are given in Table \ref{SCCs}. Finally, we chose four parameters of each observable with the largest absolute values of SCC: $U_{2}$, $T_{10}-T_{0}$, $T_{14}$ and $U_{4}-U_{2}$, where $U$ indicates wind speed, $T$ temperature, and the subscript is the height in metres. Of these four parameters, $T_{10}-T_{0}$ and $U_{2}$ are the most significant. Fig. \ref{fig:seeing and meteorological parameter statistics} shows the relationship between $T_{10}-T_{0}$, $U_{2}$ and the seeing.  

The optical turbulence can be roughly characterized by the Richardson number \citep{aristidi05}, which is defined as:
\begin{gather}
	\label{cell state}
	R_{i}=\frac{g}{\theta} \frac{(d\theta /dz)}{(dU/dz)^2}, 
\end{gather}
where $R_i$ is the Richardson number, $g$ is the gravitational constant, $d\theta /dz$ and $dU/dz$ are the potential temperature and wind speed gradients with respect to the height $z$. The potential temperature of a parcel of air is the temperature of the parcel if its pressure is adiabatically changed to a reference pressure. 
When $R_i$ is less than 0.25, optical turbulence develops. 

The potential temperature can be derived from the strength of the temperature inversion, with $R_i$ increasing (and optical turbulence decreasing) as the temperature inversion increases. Therefore, as expected, a clear anti-correlation between $T_{10}-T_{0}$ and the seeing is seen in Fig. \ref{fig:seeing and meteorological parameter statistics} (a). The absolute SCC between $T_z-T_0$ and the seeing increases until $z$ reaches 10 m. This is because the KL-DIMM was installed at 8 m,  which makes $T_{10}-T_{0}$ the best measurement of the strength of the temperature inversion for our model. $U_{4}-U_{2}$ is also selected as it measures the wind speed gradient contribution to $R_i$. However, the Richardson number is only a rough indicator of optical turbulence, and there are other parameters related to the seeing. Fig. \ref{fig:seeing and meteorological parameter statistics} (b) shows a strong correlation between $U_2$ and the seeing. Optical turbulence can be generated by air moving in a non-laminar fashion over the rough ground surface, and so it is unsurprising that the seeing is almost always larger than 1 arcsec when the wind near the ground is relatively high, $U_2 >$ 4 m/s. In summary, the parameters chosen for our model by selecting for high SCC have a reasonable physical interpretation for their effects. 

\section{MODEL FRAMEWORK}
\label{sec:model}

We adopt a LSTM-GPR (long short-term memory, gaussian process regression) model to predict the seeing from multi-layer meteorological data. The workflow of our framework is illustrated in Fig. \ref{fig:LSTM_predictor}. Our framework has two components: the seeing estimator derived from meteorological parameters using GPR and the meteorological parameter predictors using LSTM. More specifically, seeing estimation is the premise of seeing prediction. The regression model of meteorological parameters and seeing is developed using the GPR model. Subsequently, LSTM is used to predict meteorological parameters, and the predicted values of meteorological parameters are fed into the established GPR model to predict seeing. The details of these two components are presented in the subsequent parts of this section. 

\begin{table*}
        \setlength{\belowcaptionskip}{0.2cm}
	\centering
	\caption{The Spearman's rank correlation coefficients between the seeing and all meteorological parameters. $T_{h}$, $ U_{h}$, $ Wd_{h}$ are temperature, wind speed and wind direction at heights of $h$ m respectively,  
	$Sl$ represents the solar altitude, $Rh_{2}$ represents the relative humidity, $Ht_{2}$ is the temperature from the sensor of relative humidity at 2 m,  
	$Ap_{2}$ represents the air pressure at 2 m, and $Apin$ represents the air pressure in the electronic box (at 2 m). The parameters we used in our model are in bold.
  }
\begin{tabular}{cccccccccccc}
	\hline
	\multicolumn{2}{c}{\begin{tabular}[c]{@{}c@{}}Temperature\\  $T_{h}$\end{tabular}} & \multicolumn{2}{c}{\begin{tabular}[c]{@{}c@{}}TI\\     $T_{h}$-$T_{0}$\end{tabular}} & \multicolumn{2}{c}{\begin{tabular}[c]{@{}c@{}}Wind speed\\ $U_{h}$\end{tabular}} & \multicolumn{2}{c}{\begin{tabular}[c]{@{}c@{}}Wind direction   \\ $Wd_{h}$\end{tabular}} & \multicolumn{2}{c}{\begin{tabular}[c]{@{}c@{}}Wind speed gradient  \\ $U_{h}-U_{2}$\end{tabular}} & \multicolumn{2}{c}{Other parameters} \\ \hline
	$T_{-1}$                                  & 0.12                                   & $T_{-1}- T_{0}$                               & -0.08                              & \textbf{$U_{2}$}                         & \textbf{0.76}                         & $Wd_{2}$                                      & 0.07                                    & \textbf{$U_{4}-U_{2}$}                             & \textbf{-0.33}                             & $Sl$           & -0.10          \\
	$T_{0}$                                   & 0.30                                   & $T_{1}- T_{0}$                                & -0.17                              & $U_{4}$                                  & 0.69                                  & $Wd_{4}$                                      & 0.03                                    & $U_{6}-U_{2}$                                      & -0.30                                      & $Rh_{2}$           & -0.08             \\
	$T_{1}$                                   & 0.09                                   & $T_{2}- T_{0}$                                & -0.37                              & $U_{6}$                                  & 0.63                                  & $Wd_{6}$                                      & 0.04                                    & $U_{8}-U_{2}$                                      & -0.14                                      & $Ap_{2}$          & -0.06            \\
	$T_{2}$                                   & -0.03                                  & $T_{4}- T_{0}$                                & -0.59                              & $U_{8}$                                  & 0.64                                  & $Wd_{8}$                                      & 0.04                                    & $U_{10}-U_{2}$                                      & -0.02                                      & $Apin$              & -0.10             \\
	$T_{4}$                                   & -0.26                                  & $T_{6}- T_{0}$                                & -0.69                              & $U_{10}$                                 & 0.64                                  & $Wd_{10}$                                     & 0.01                                    & $U_{12}-U_{2}$                                     & -0.19                                      & $Ht_{2}$           & -0.04            \\
	$T_{6}$                                   & -0.41                                  & $T_{8}- T_{0}$                                & -0.68                              & $U_{12}$                                 & 0.66                                  & $Wd_{12}$                                     & 0.06                                    & $U_{14}-U_{2}$                                     & -0.32                                      & -                 & -                \\
	$T_{8}$                                   & -0.48                                  & \textbf{$T_{10}- T_{0}$}                     & \textbf{-0.70}                     & $U_{14}$                                  & 0.71                                  & $Wd_{14}$                                     & 0.08                                     & -                                                & -                                          & -                 & -                \\
	$T_{10}$                                  & -0.52                                  & $T_{12}- T_{0}$                               & -0.69                              & -                                        & -                                     & -                                             & -                                       & -                                                  & -                                          & -                 & -                \\
	{$T_{12}$}                                & -0.53                                  & $T_{14}- T_{0}$                               & -0.69                              & -                                       & -                                     & -                                             & -                                       & -                                                  & -                                          & -                 & -                \\ 
	\textbf{$T_{14}$}                         & \textbf{-0.56}                         & -                                             & -                                  & -                                       & -                                     & -                                             & -                                       & -                                                  & -                                          & -                 & -                \\ \hline
	\label{SCCs}
\end{tabular}
\end{table*}

\subsection{Seeing estimation method} 
\label{Seeing estimation method} 

GPR is a powerful tool to approximate nonlinear and linear functions \citep{Rasmussen06,Bu18}. The significant advantage of GPR over neural network (NN) algorithms is that GPR allows the prediction uncertainty to be quantified in a principled way \citep{Soares22}. A Gaussian process can be viewed as an extension of a multivariate Gaussian distribution to infinite dimensions. The dataset for seeing estimator is $D=\{x_{n},y_{n}\}_{n=1}^{N}$ where the inputs $x=\{x_{n}\}_{n=1}^{N}$ are meteorologic values and targets $y=\{y_{n}\}_{n=1}^{N}$ are seeing values. We assume that the seeing data have been created by a latent function $y=f(x)$. All values of $f(x)$ at any location $x$ are determined by the mean function $m(x)$ and the covariance function $K(x,x^{'})$: 
\begin{gather}
f(x)=GP(m(x),K(x,x^{'}))),
\end{gather}
where $m(x)=[m(x_1),m(x_2),...,m(x_n)]$ is the vector of mean values, and $K$ is the $n \times n$ covariance matrix with $(i, j)$, the element $K_{ij} = k(x_i, x_j)$.
In the Gaussian process, the covariance function expresses the similarity between the features of the samples. Therefore, this method is suitable for modeling time series data, such as seeing and meteorological data. 
The covariance function $K(x_{i},x_{j})$ can be defined by various kernel functions and parameterized in terms of the kernel parameters in vector $\theta$. 

Here, we adopt
a commonly used covariance function, the Rational Quadratic kernel,  
given by:
\begin{gather}
k(x_{i},x_{j}|\theta )=\theta _{1}(1+\frac{(x_{i}-x_{j})^{T}(x_{i}-x_{j})}{2\alpha \theta_{2} }),
\end{gather}
where the kernel parameters $\theta_{i}$ can be estimated by maximizing the marginal likelihood of the joint Gaussian distribution.   
For a more detailed introduction to GPR, readers can refer to \citet{Rasmussen06}.

\subsection{Seeing prediction method} 
\label{Seeing prediction method} 

Because of their excellent non-linear fitting abilities, artificial neural networks are widely used in weather forecasting and related areas \citep{French92,Gill10,Karevan20}. One family of ANNs, named recurrent neural networks (RNNs), are specialized for processing sequential data \citep{Rumelhart86}.

\begin{figure}
	\centering  
	\includegraphics[width=\columnwidth]{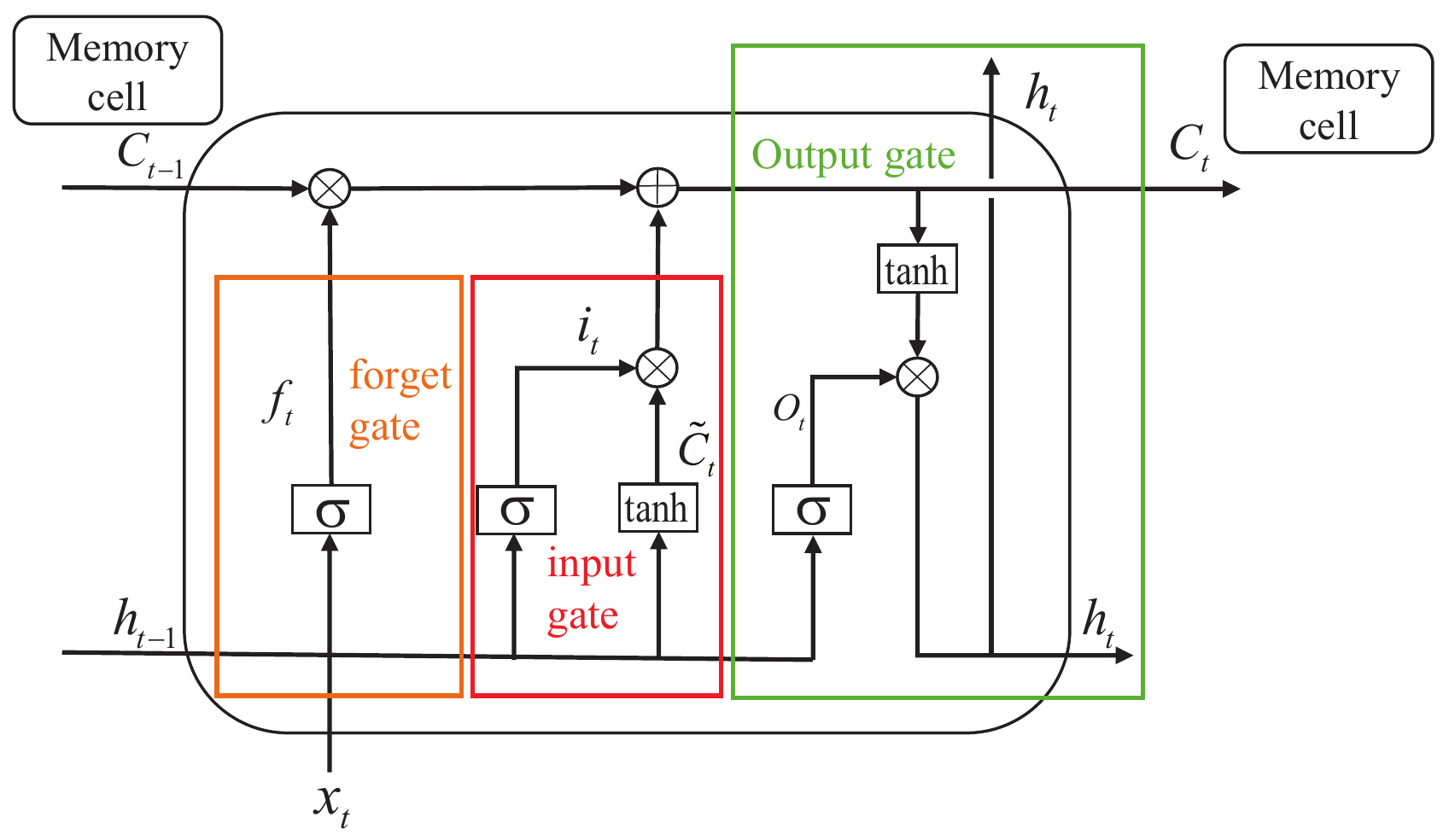}
	\caption{The schematic diagram of the LSTM unit. The rows represent the flow of information, starting from the inputs 
	(previous time-step hidden state $h_{t-1}$, memory cell state $c_{t-1}$ and current input sequence $x_{t}$) to the new hidden state $h_{t}$ and memory cell state $c_{t}$. }
	\label{fig: LSMT_cell}
\end{figure}

However, RNNs cannot learn long-term temporal correlations due to vanishing and exploding gradient problems. To overcome these issues, the LSTM network has been proposed \citep{Hochreiter97}. The LSTM network is a modified RNN algorithm where neurons can maintain memory in their channels to alleviate the problem of vanishing and exploding gradients \citep{Salman18}. The meteorological data has strong time-dependency features, while the LSTM network can effectively capture the relationships and dependencies for time series. Therefore, the LSTM network provides a promising method in seeing prediction. The key to the LSTM network is the cell state; the structure of the LSTM is shown in Fig. \ref{fig: LSMT_cell}. It consists of one memory cell to store the past information about the network and three gates (Input, Forget, and Output) to control the flow of information to the cell state. Typically, the input gate controls the amount of input information that should be saved at the current time, the forget gate controls the amount of information that should be forgotten, and the output gate controls when the LSTM cell should output the value. The three gates at time $t$ are calculated as follows: 
\begin{gather}
	\label{input_gate}
	i_{t}=\sigma(W_{i}\circ [h_{t-1},x_{t}]+b_{i}),\\
	f_{t}=\sigma(W_{f}\circ [h_{t-1},x_{t}]+b_{f}),\\
	o_{t}=\sigma(W_{o}\circ [h_{t-1},x_{t}]+b_{o}),
\end{gather}
where $x_{t}$ is the input vector at the time $t$, $h_{t-1}$ is the previous hidden state, $W$ and $b$ are the recurrent weight and bias respectively, 
$\sigma$ is the sigmoid activation function and $\circ$ denotes matrix multiplication.  

The memory cell state $c_{t}$ and the LSTM hidden state $h_{t}$ are updated as follows:  
\begin{gather}
	\label{cell state}
	c_{t}=f_{t}\otimes c_{t-1}+i_{t}\otimes \tilde{c}_{t},\\
	h_{t}=o_{t}\otimes\text{tanh}(c_{t}),
\end{gather}
where $c_{t-1}$ and $h_{t-1}$ represent the previous memory cell state and hidden state respectively. The operation $\otimes $ is the Hadamard (Element-Wise) product. $\tilde{c}_{t}$ is candidate value, which is calculated by:
\begin{gather}
	\label{cell state}
	\tilde{c}_{t}=\text{tanh}(W_c\circ [h_{t-1},x_{t}]+b_{c}),
\end{gather}

\begin{figure}
	\includegraphics[width=\columnwidth]{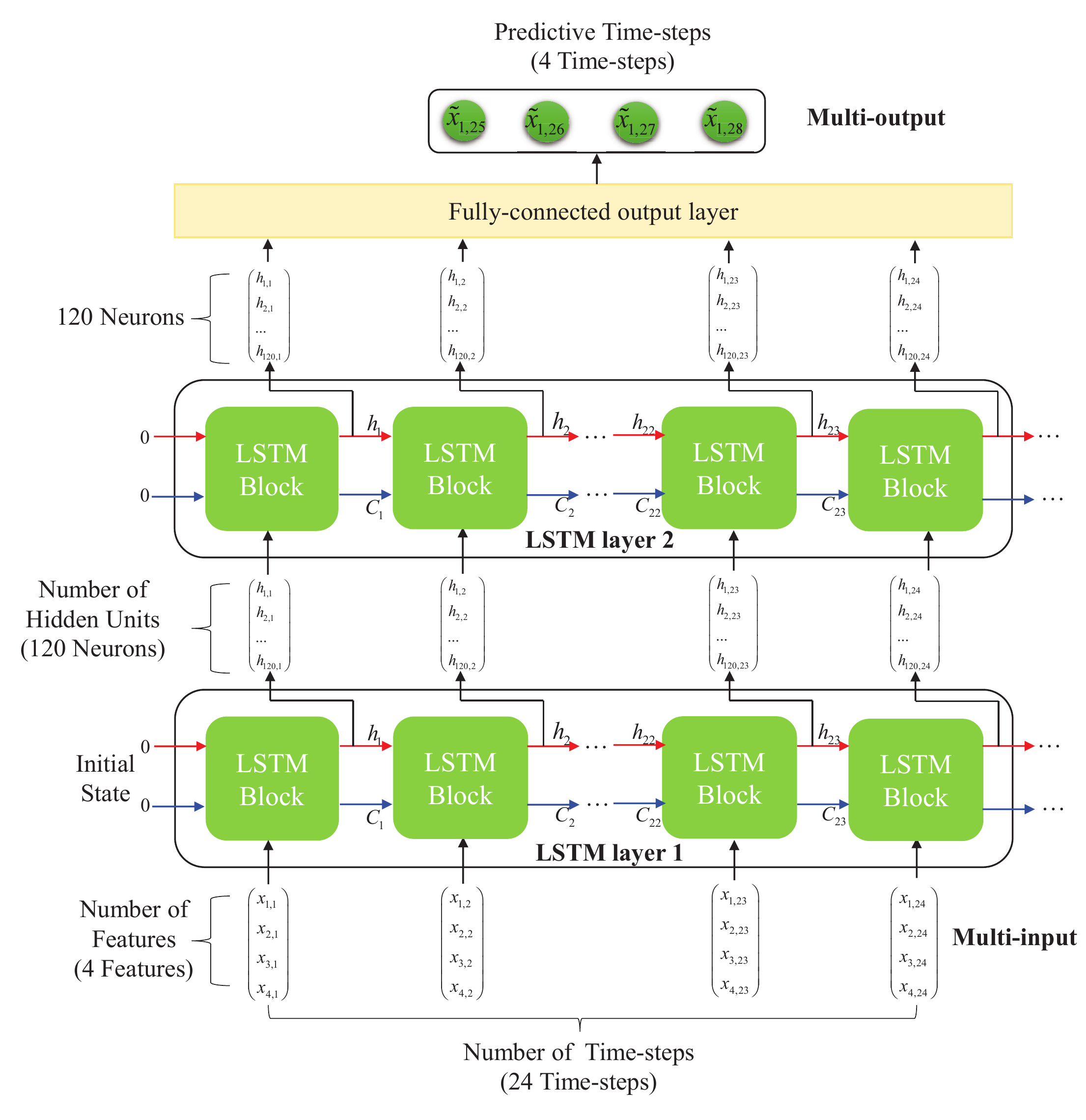}
	\caption{Diagram of the proposed LSTM neural network. }
	\label{fig: LSTM_structure}
\end{figure}

The following parameters are used for LSTM in our study: the data set is divided into a training set
and a testing set in the ratio of 8:2, the activation function is tanh, the loss function is mean squared error, 
the optimizer is Adam, the number of features is 4 (temperature, TI, wind speed, and wind speed gradient), and the number of neurons in the output layer is 4. 
To achieve better performance of the LSTM model, we tuned three key hyper-parameters that determine the 
physical capacity of the network, including the input time-sequence length, the number of neurons in each hidden layer, and the number of hidden layers. 
For each set of these hyper-parameters are chosen, the model is recompiled, re-initialized, and retrained. 
Finally, these three hyper-parameters are configured such that 
the input time-sequence is 120 minutes (i.e., time-steps is 24), the number of layers is two, and the number of neurons is 120 for each hidden layer. 
The detailed network structure is shown in Fig. \ref{fig: LSTM_structure}. 

\section{RESULT}
\label{sec:res}
To reduce the effect of different probability distributions in the training and testing sets, 
we used the 5-fold cross-validation (CV) method to train and evaluate our model, a common technique used in machine learning \citep{He15,Waris19}. 
The 5-fold CV partitions the data randomly into five mutually exclusive folds, with four subsets for the training and the remaining subset for testing. This method preserves the consistency of each subset distribution as much as possible. 
We used the RMSE and coefficient of determination $R^{2}$ to evaluate the performance of the model on the test data:  
\begin{gather}
	\label{function_RMSE_R2}
	RMSE=\sqrt{\frac{1}{n}\sum_{i=1}^{n}(y_{i}-y_{i}^{\ast })^{2}}, \\
	R^{2}=1-\frac{\sum_{i=1}^{n} (y_{i}^{\ast}-\bar{y})^2}{\sum_{i=1}^{n} (y_{i}-\bar{y} )^2},
\end{gather}
where $\bar{y}$ is the mean of the observed values, $y_{i}^{\ast }$ is the predicted or estimated value, and $n$ is the number of samples.

In the following section, we apply the LSTM-GPR model to the meteorological dataset and estimate the future seeing. Firstly, we evaluate the performance of the GPR in estimating the seeing, then we analyse the seeing prediction ability of LSTM-GPR. Finally, we compare the computation speed of WRF model and our proposed LSTM-GPR model.

\subsection{Seeing estimation performance}
\label{Seeing estimation result}

\begin{figure}
	\includegraphics[width=\columnwidth]{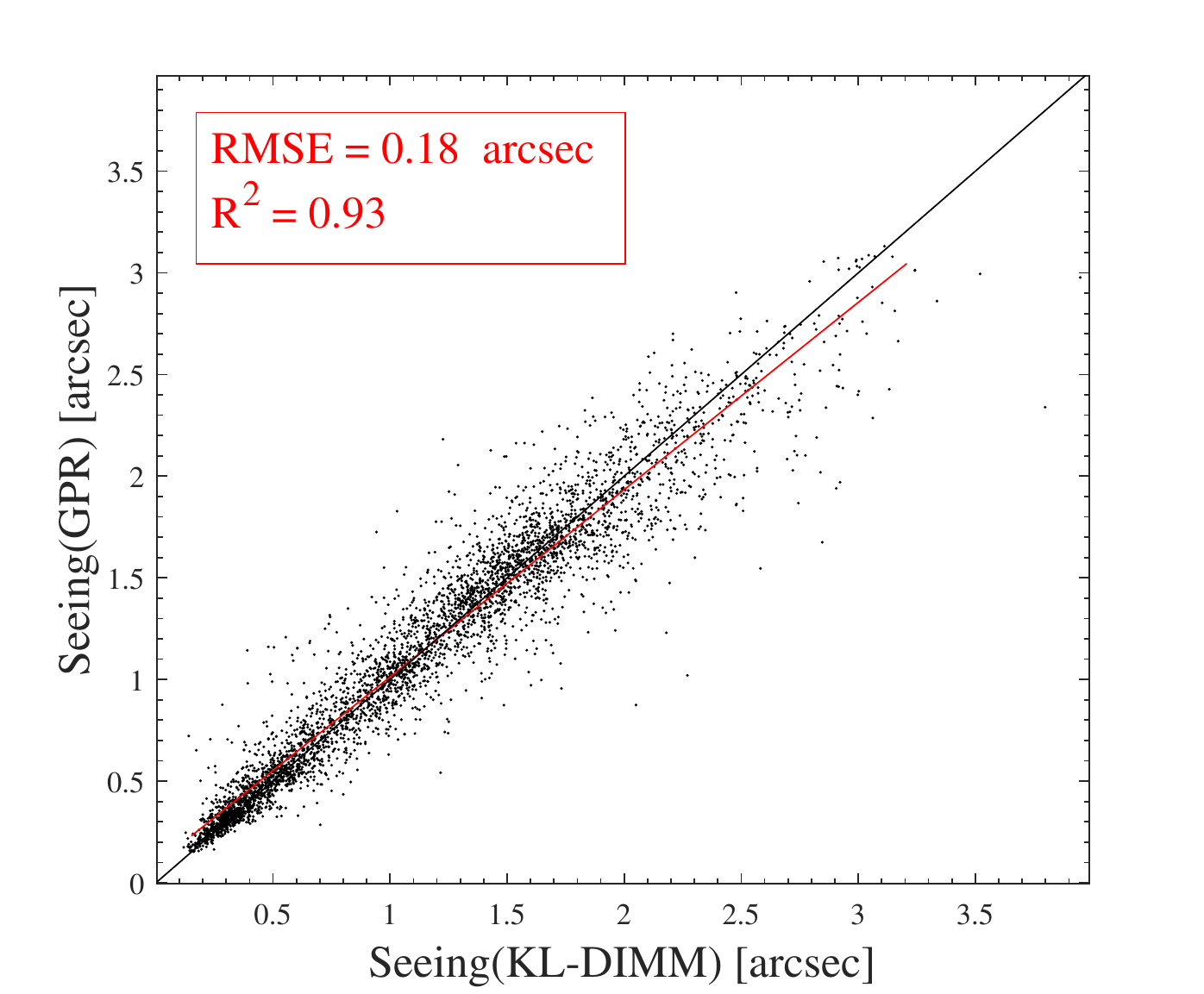}
	\caption{Correlation between seeing measured by KL-DIMM (abscissa) and estimated by the GPR model (ordinate).The black line follows y=x.}
	\label{fig:seeing_estimate}
\end{figure}

Fig. \ref{fig:seeing_estimate} shows the scatter plot between the seeing estimated by GPR and the measurements from the test data set. The regression coefficient of the linear fitting model is 0.91 (red line). It can be seen from this graph that the RMSE of the seeing estimate is 0.18 arcsec, and $R^2$ is 0.93. The estimated results (black points) lie closely along the diagonal line, showing the high consistency between the seeing estimation and measurements. 

\begin{figure*}
	\centering  
	\subfigure[]{
		\includegraphics[width=0.29\linewidth]{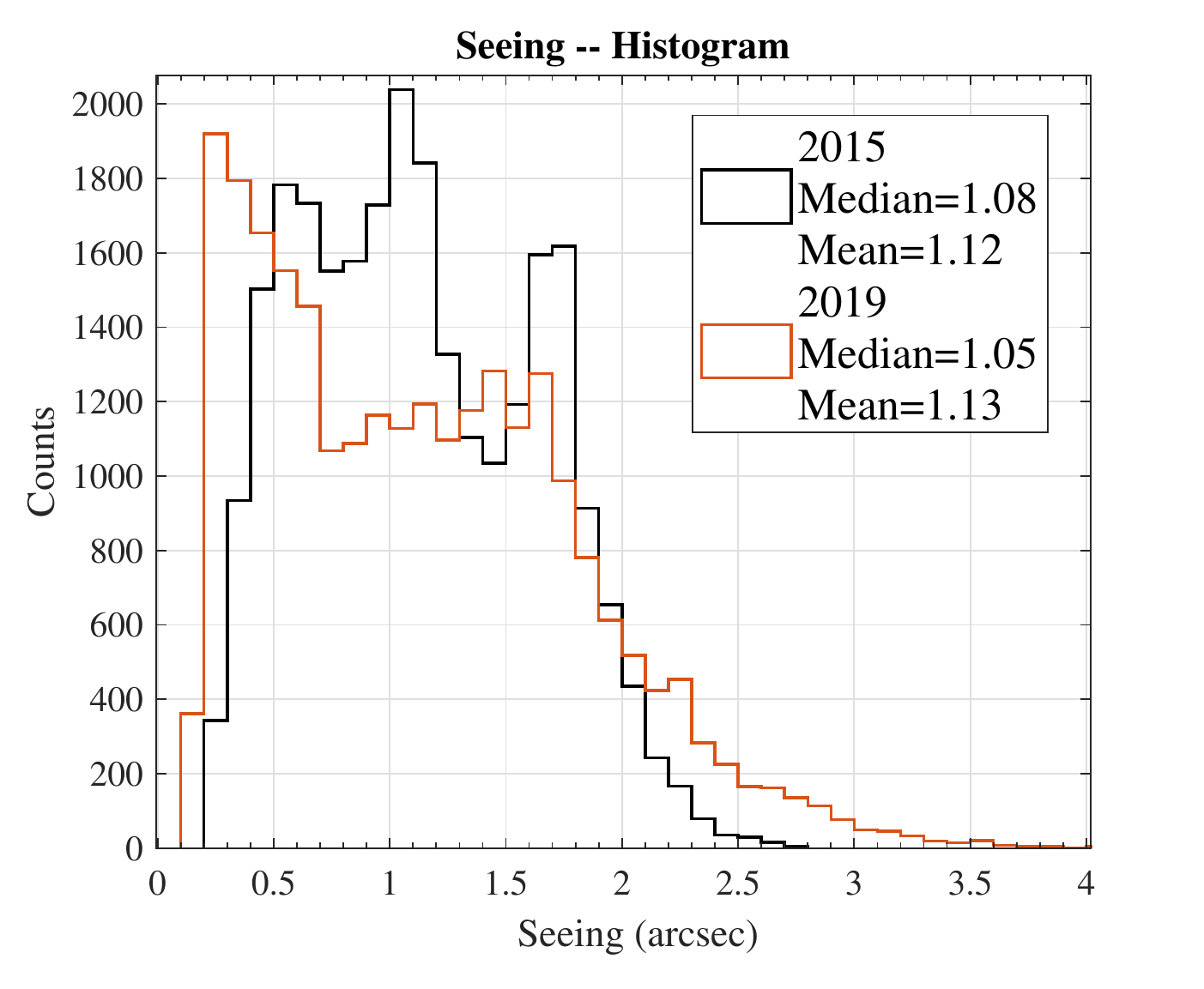}}
	\subfigure[]{
		\includegraphics[width=0.29\linewidth]{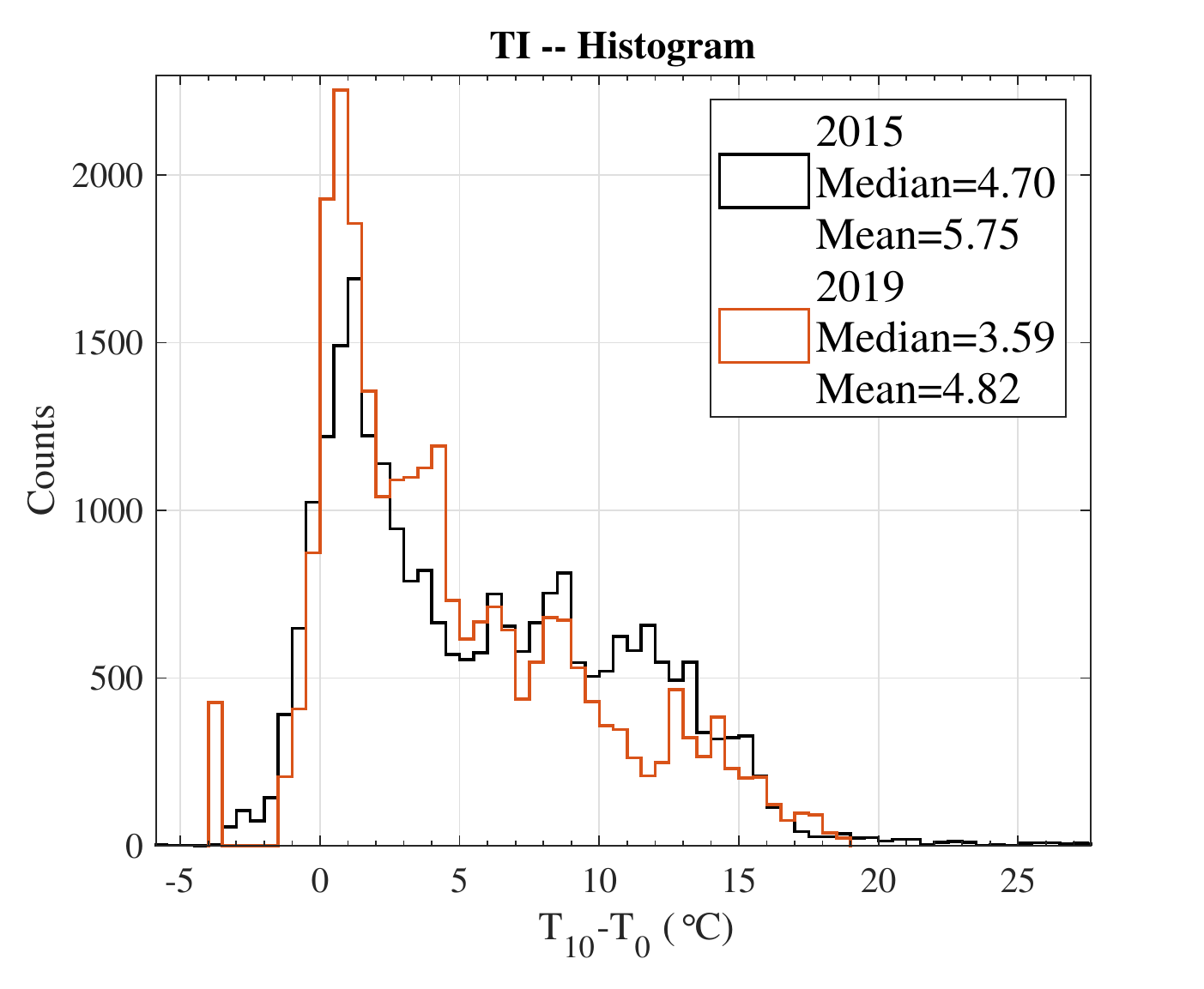}}
	\subfigure[]{
		\includegraphics[width=0.29\linewidth]{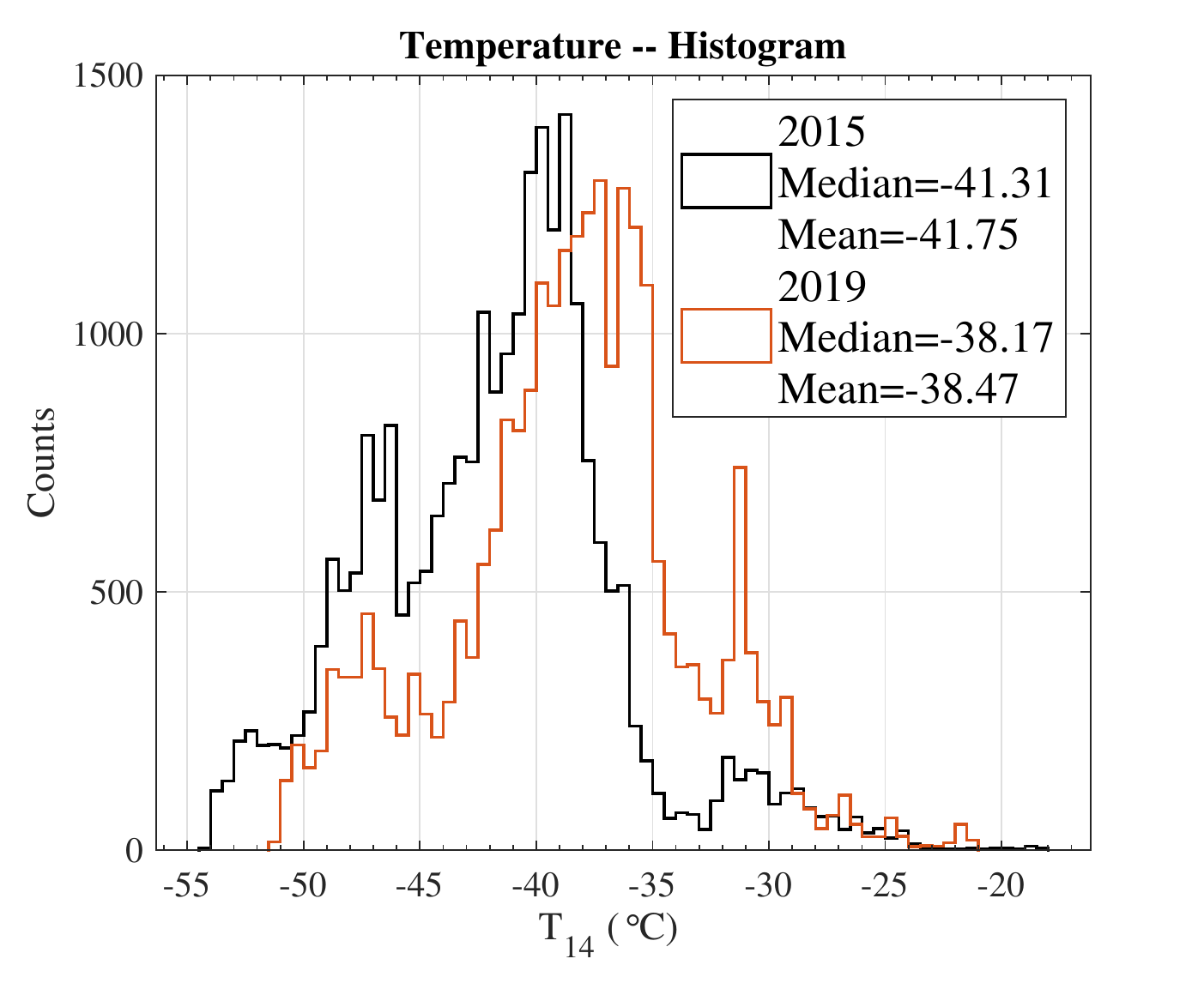}}
	\subfigure[]{
		\includegraphics[width=0.29\linewidth]{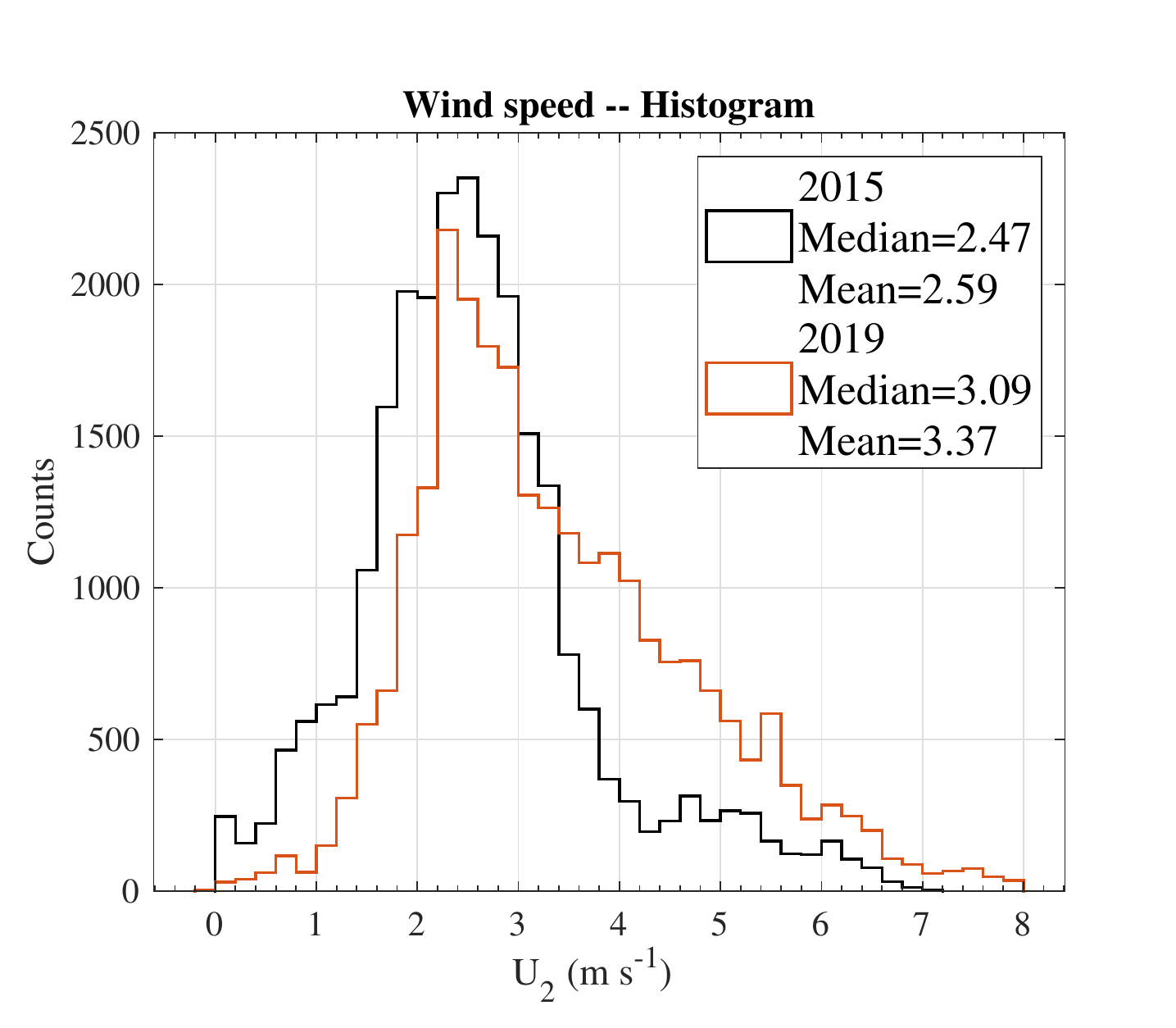}}
	\subfigure[]{
		\includegraphics[width=0.29\linewidth]{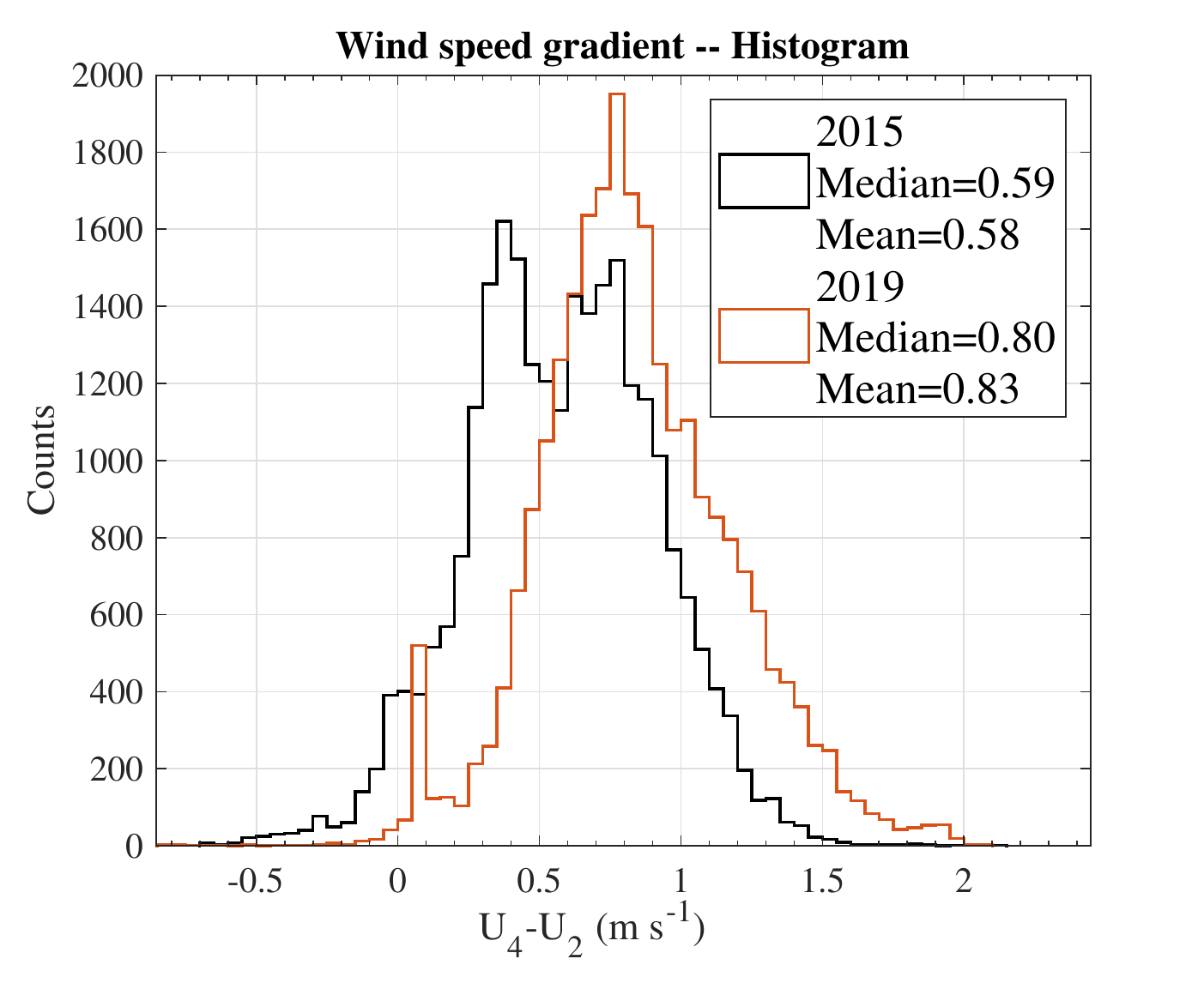}}

	\caption{Histogram of the seeing, $T_{10}-T_{0}$, $T_{14}$, $U_{2}$ and $U_{4}-U_{2}$ in 2015 (black line) and 2019 (orange line). 
	(a) Comparison of the day-time seeing estimated by the GPR in 2015 and measurements in 2019. 
	(b) Day-time $T_{10}-T_{0}$. 
	(c) Day-time $T_{14}$. 
	(d) Day-time $U_{2}$. 
	(e) Day-time $U_{4}-U_{2}$.  
	}
	\label{fig:seeing statistics}
\end{figure*}

We then used our seeing estimate model to calculate the day-time seeing using the meteorological data from KLAWS in 2015 from February 11 to March 7 \citep{Hu19}, which is the same period for which we have observational seeing and meteorological data in 2019. 
While we have no seeing observations from 2015, we can use our model to compare the estimated seeing with the observations from 2019. 
The day-time median and average seeing in 2015 derived by the GPR model are 1.08 and 1.12 arcsec, respectively, close to 1.05 and 1.13 arcsec as measured by KL-DIMM in 2019. 
The distributions of estimated day-time seeing in 2015 and the observed seeing in 2019 are shown in Fig. \ref{fig:seeing statistics} (a). 
As listed in Table \ref{SCCs}, $T_{10}-T_{0}$, $T_{14}$, $U_{2}$ and $U_{4}-U_{2}$ are the four most significant parameters that affect the seeing. 
Although the distributions of day-time $T_{10}-T_0$ and $U_{2}$ in 2015 and 2019 are similar, 
the peak location of the distribution of day-time $T_{14}$ and $U_{4}-U_{2}$ in 2019 are significantly larger than that in 2015 (Fig. \ref{fig:seeing statistics} (b)-(e)), which could explain the slightly larger median day-time seeing in 2015, and is likely due to natural variations in the atmospheric conditions.

\subsection{Seeing prediction performance}
\label{Seeing prediction result} 

The LSTM model is adopted to build the weather forecast model and predict four meteorological parameters for the next 20 minutes. 
The meteorological data used in this paper was collected from 2019 February 11 to March 7. 
Eighty per cent of the data were used (from February 11 to March 2) for network training, while the rest (from March 2 to 7) were used for testing. 

\begin{figure*}
	\centering  
	\subfigure{
		\includegraphics[width=0.4\linewidth]{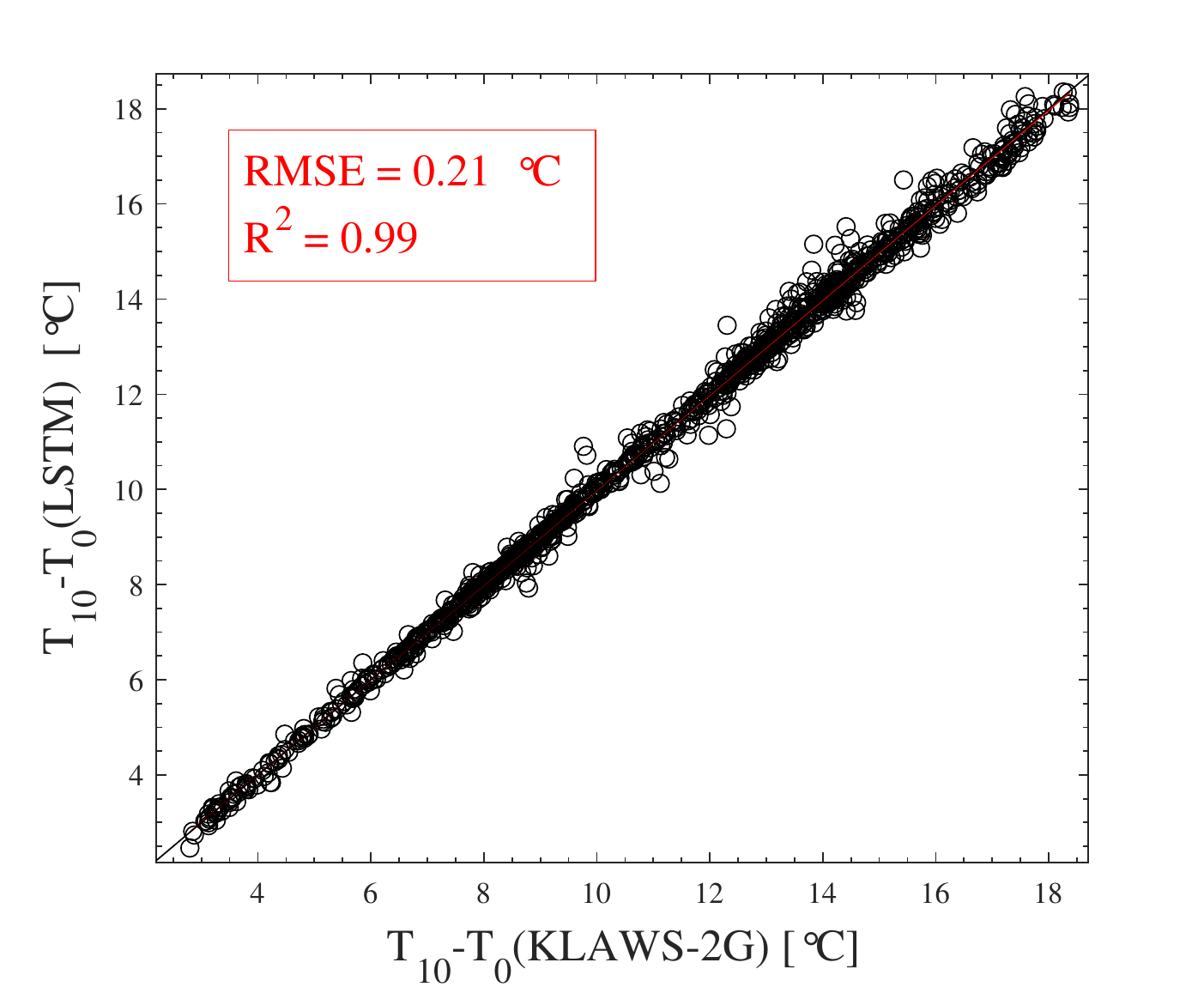}}
	\subfigure{
		\includegraphics[width=0.4\linewidth]{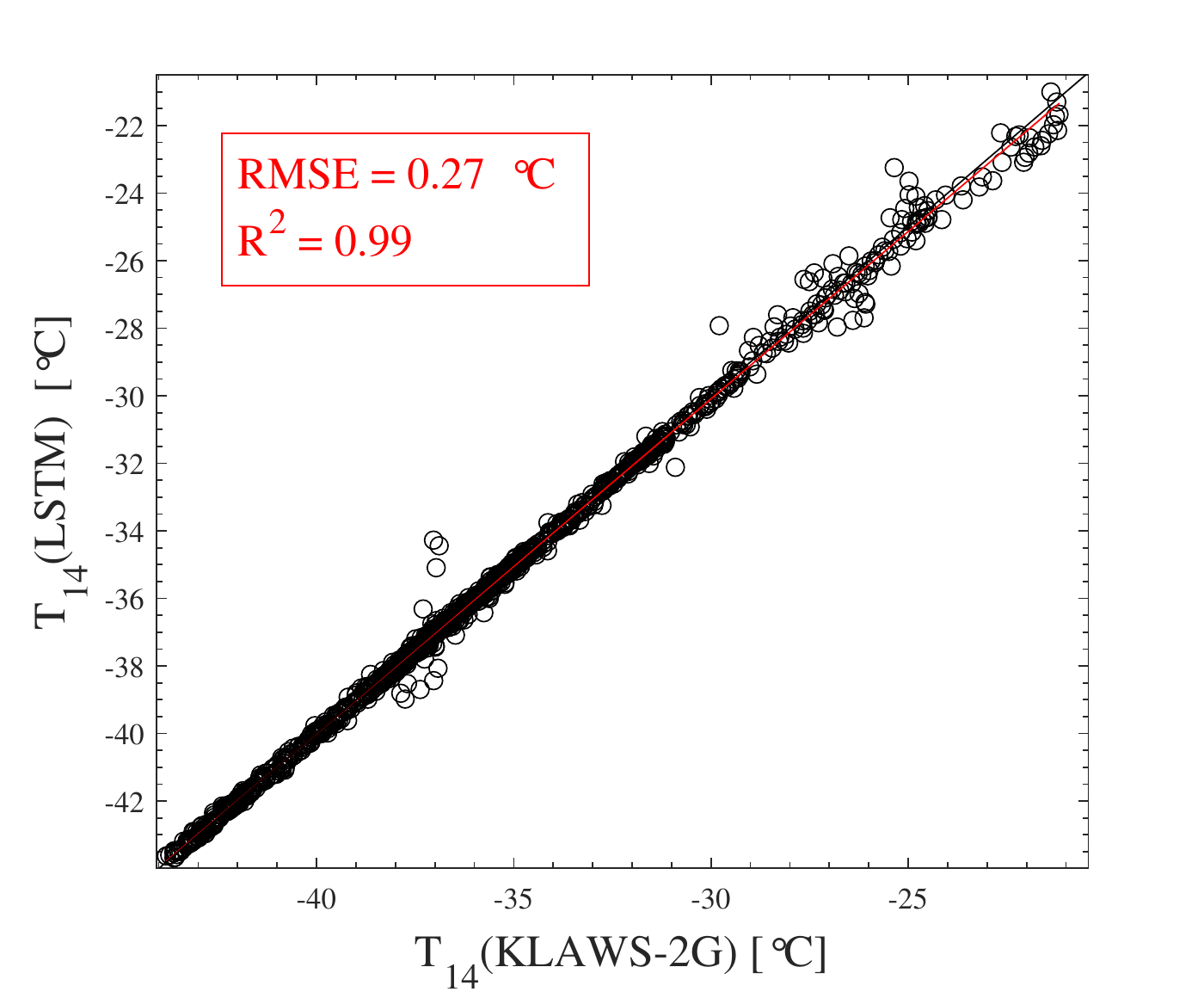}}
	\quad
	\subfigure{
		\includegraphics[width=0.4\linewidth]{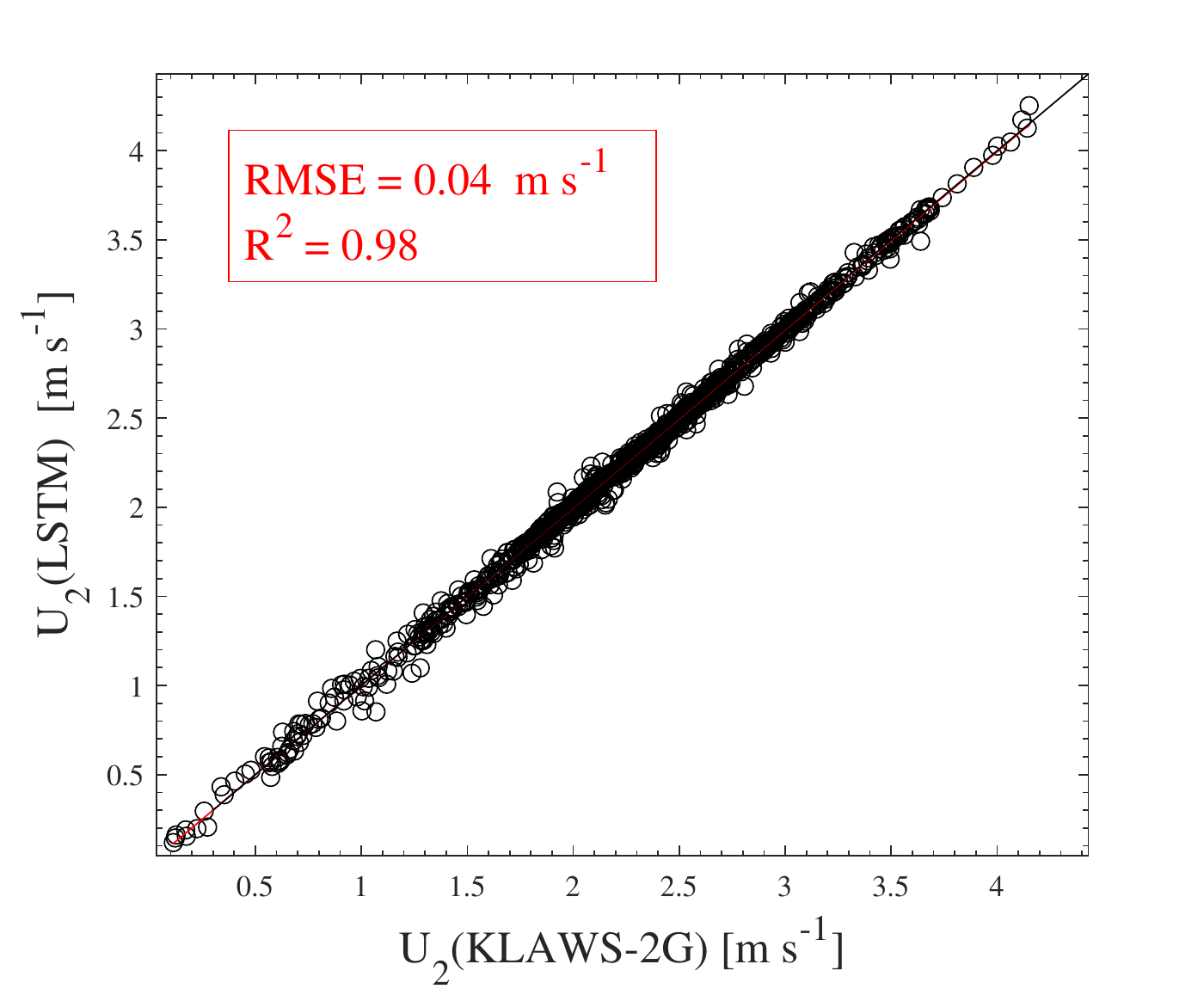}}
	\subfigure{
		\includegraphics[width=0.4\linewidth]{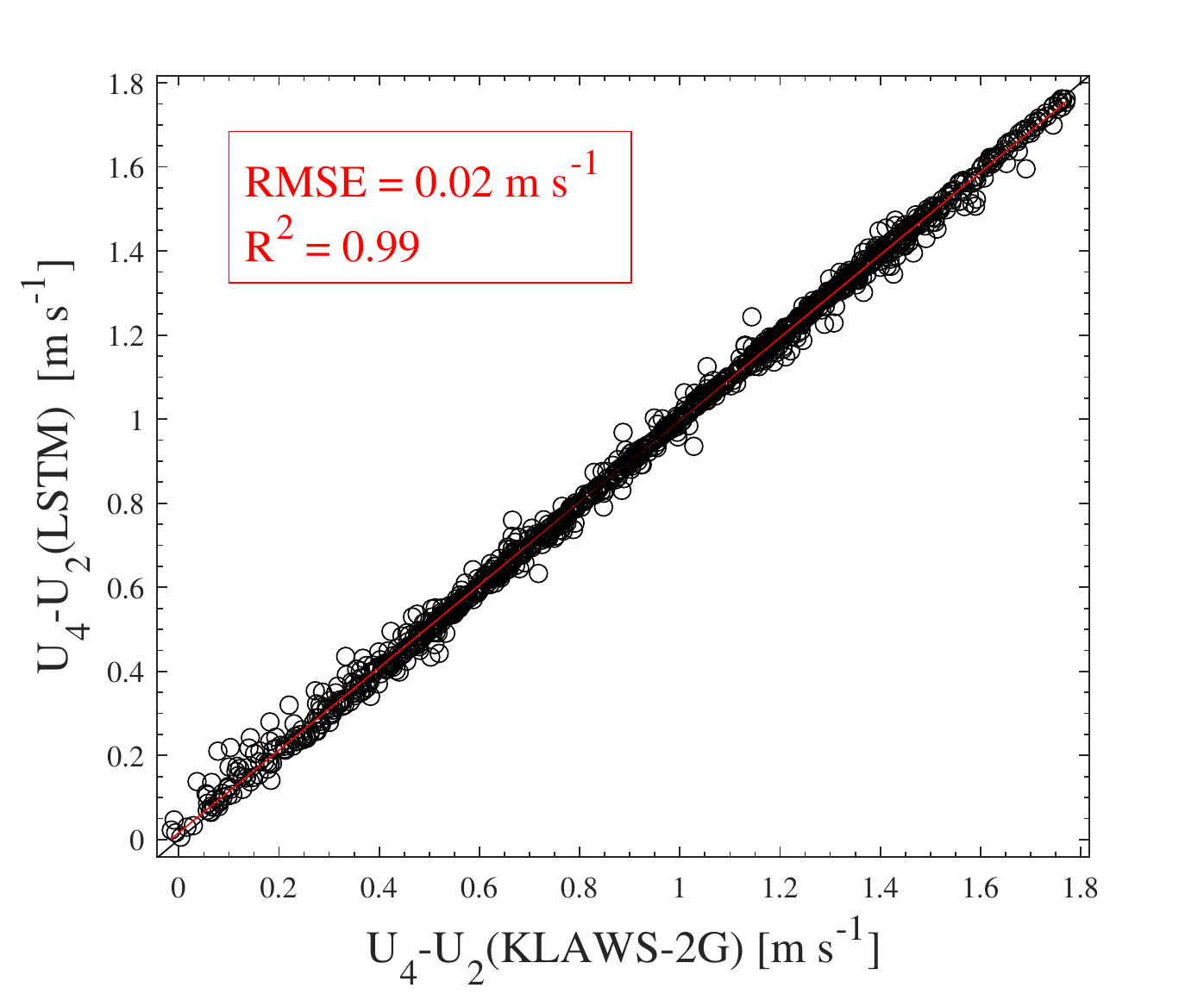}}
	\caption{Correlation between meteorological parameters collected by KLAWS-2G (abscissa) and predicted by the LSTM model (ordinate). From the top left to the bottom right: TI, temperature, wind speed, and wind speed gradient.  }
	\label{fig:weather scatter}
\end{figure*}

Fig. \ref{fig:weather scatter} shows the comparison of all the meteorological parameters as measured by KLAWS-2G and predicted by LSTM.  
It can be seen from this figure that the RMSE is 0.21 $^{\circ}$C for the $T_{10}-T_{0}$, 0.27 $^{\circ}$C for the $T_{14}$, 0.04 m $\rm s^{-1}$ for the $U_{2}$ and 0.02 m $\rm s^{-1}$ for the $U_{4}-U_{2}$ respectively. 
The corresponding $R^2$ are 0.99, 0.99, 0.98 and 0.99. 
These results give good confidence that LSTM is able to forecast the meteorological parameters.

\begin{figure*}
	\centering  
	\includegraphics[width=0.85\textwidth]{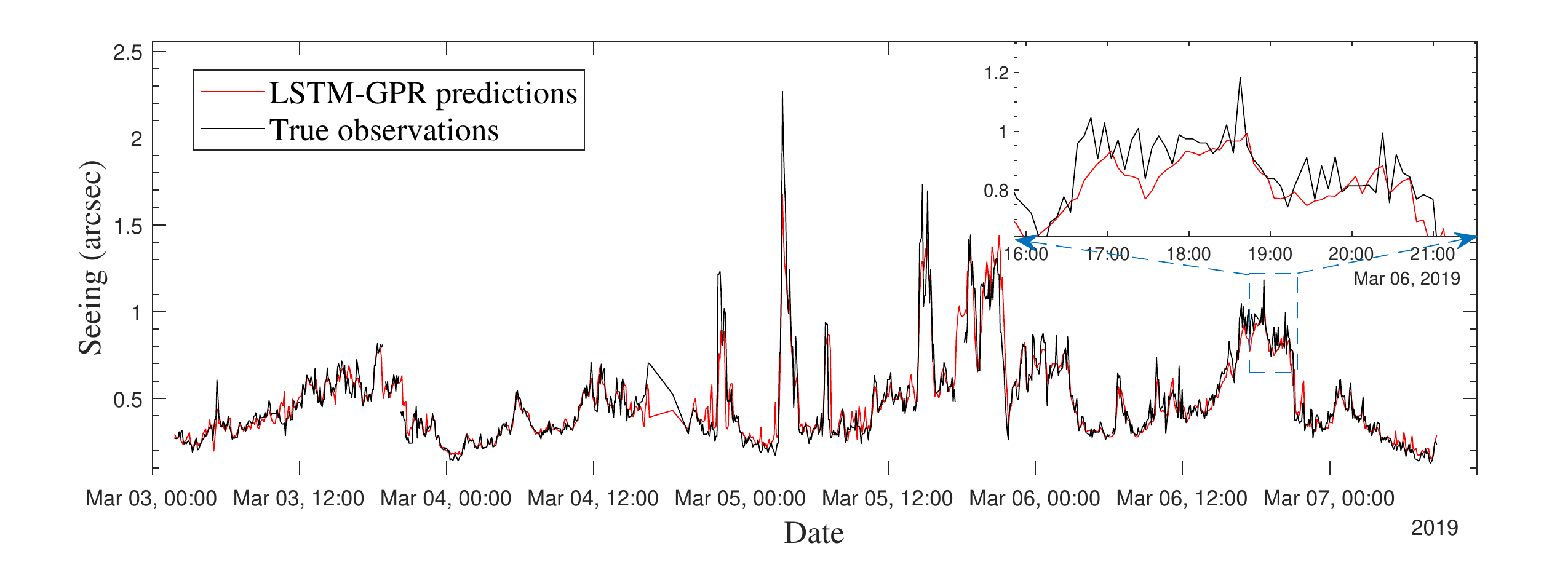}
	\caption{Seeing prediction results at 20 minutes ahead.}
	\label{fig: seeing}
\end{figure*}

\begin{figure}
	\centering
	\subfigure
	{
	 \begin{minipage}{8cm}
	  \includegraphics[width=\columnwidth]{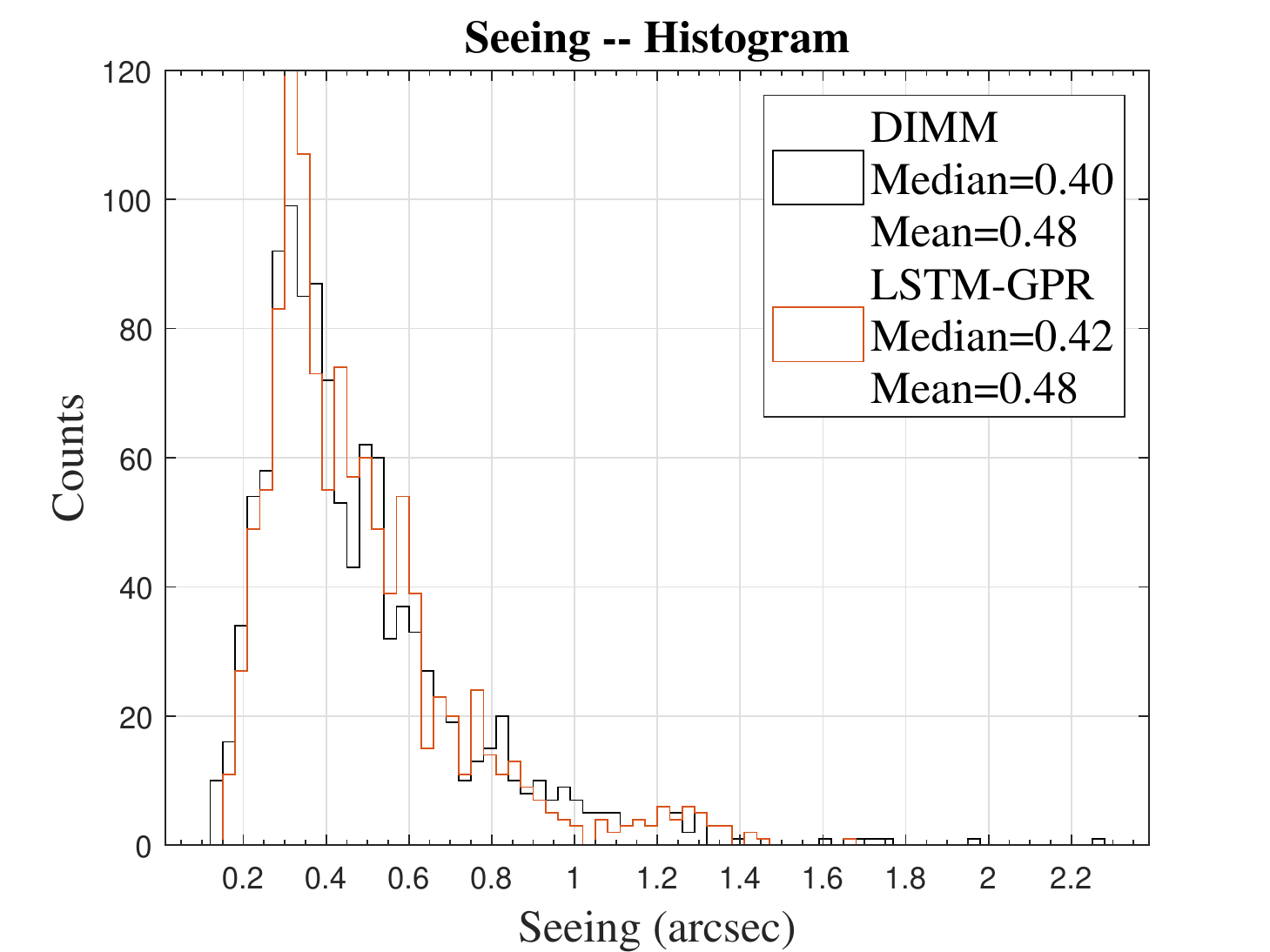}
	 \end{minipage}
	}
	   \caption{ Histogram of the seeing measured by the KL-DIMM (black line) and predicted by the LSTM-GPR model (orange line) in 2019.}
	   \label{fig:seeing predict_true statistics}
\end{figure}

The meteorological parameters predicted by LSTM were fed into the GPR model to estimate the seeing 20 minutes into the future. We compared the predicted results with actual observations from 2019 March 2 to 7 in Fig. \ref{fig: seeing}. 
The RMSE of the seeing prediction is 0.12 arcsec, and $R^2$ is 0.86. A zoomed-in view of the predicted results for a 6 hour period is shown in the inset of Fig. \ref{fig: seeing}. As can be seen from this figure, the predicted results and actual observations are in good agreement. In addition, we have plotted the distributions of the true and predicted seeing in Fig. \ref{fig:seeing predict_true statistics}, and the agreement is good.

Finally, we have compared our model with the persistence, the simplest method that is commonly used in forecasting studies \citep{Taylor02,Giordano13}. The principle of the persistence is simply that the last observed data is used as the next prediction, i.e., 
$F_{t+k}=y_{t}$, where $y_{t}$ is the latest measured value and $F_{t+k}$ is the future forecast value, and $k$ is the time-step of the forecast. The detailed results are shown in Table \ref{tab: LSTM and baseline model prediction}. It can be noticed from this table that the RMSE for both meteorological and seeing data predicted by our proposed model is lower than the persistence. Compared to the persistence, the RMSE of seeing predicted by the LSTM-GPR is reduced by 37 per cent. The $R^2$ of seeing predicted by the LSTM-GPR model is higher than the persistence.

\begin{table}
\setlength{\belowcaptionskip}{0.2cm}
\caption[]{Performance comparison of the LSTM-GPR model and the persistence on the testing set. }
\label{tab: LSTM and baseline model prediction}
\resizebox{\linewidth}{!}{
 \begin{tabular}{ccccccccccccc}
  \hline\noalign{\smallskip}
    \multirow{2}{*}{Predicted parameter} & 
    \multicolumn{2}{c}{LSTM-GPR model} & 
    \multicolumn{2}{c}{Persistence} \\ & RMSE     & $R^{2}$ & RMSE      & $R^{2}$ \\ 
  \hline\noalign{\smallskip}
$T_{10}-T_{0}$ & 0.21 $^{\circ}$C   & 0.99 & 0.93 $^{\circ}$C & 0.94 \\ $T_{14}$ & 0.27 $^{\circ}$C & 0.99 & 0.86 $^{\circ}$C & 0.97 \\ $U_{2}$ & 0.04 m $\rm s^{-1}$ & 0.98 & 0.25 m $\rm s^{-1}$  & 0.87 \\ $U_{4}-U_{2}$ & 0.02 m $\rm s^{-1}$ & 0.99 & 0.13 m $\rm s^{-1}$  & 0.89 \\ Seeing & 0.12 arcsec & 0.86 & 0.19 arcsec  & 0.62 \\
\noalign{\smallskip}\hline
\end{tabular}
}
\end{table}

\begin{figure}
\includegraphics[width=\columnwidth]{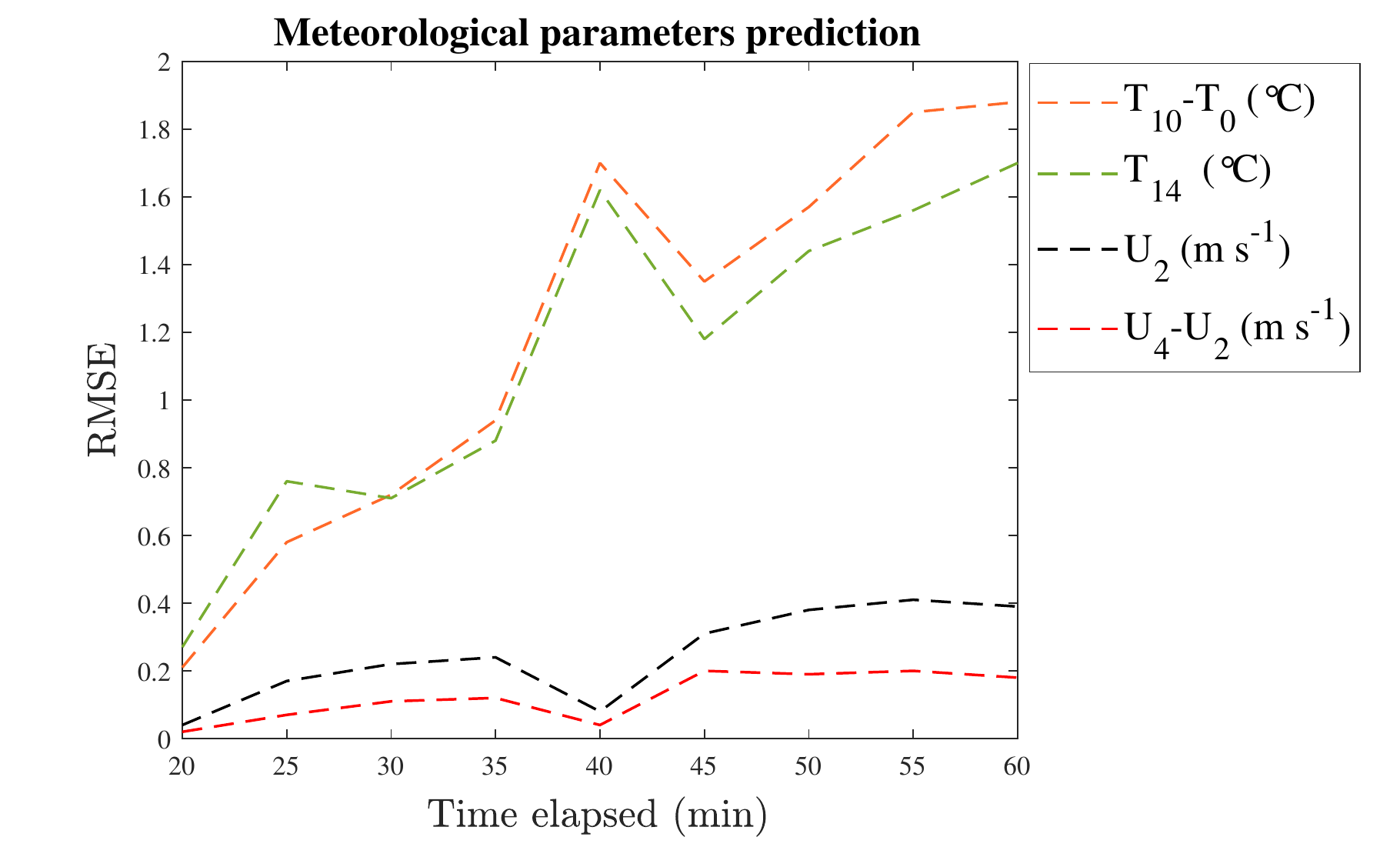}
\caption{Prediction error of meteorological parameters on different forecasting horizons.}
\label{fig: time_elapsed}
\end{figure}

To evaluate the performance of the LSTM-GPR model over different time horizons, we predicted the meteorological parameters and seeing from 20 to 60 minutes.
Fig. \ref{fig: time_elapsed} shows the RMSE of the predicted meteorological parameters. The RMSE increases as the time horizons increase, although the curve of RMSE is not linear with time. 
Table \ref{tab: different time-step} compares the RMSE and $R^2$ of the predicted seeing between the LSTM-GPR model and the persistence on different time horizons. 
It can be seen from this table that our proposed method provides a better result than the persistence, with a reduction in RMSE ranging from 8 per cent to 37 per cent over the 20 to 60 minutes prediction period. The performance of the LSTM-GPR model is better than the persistence.
We also compared the result of our method with the classical mesoscale non-hydrostatic numerical weather prediction system. \citet{Yang21} use the Polar-WRF method to predict the day-time seeing at Dome A, the RMSE between measurements and forecasts is 0.36 arcsec over 24 hours timescale, while the RMSE of our proposed method is 0.24 arcsec over a one-hour timescale. But the purpose of this study is to real-time guide the telescope. Therefore, the short-term forecast time is more important.

\begin{table}
\setlength{\belowcaptionskip}{0.2cm}
\caption[]{Performance comparison of the LSTM-GPR model and the persistence for seeing prediction on different time horizons.}
\label{tab: different time-step}
\small
\resizebox{\linewidth}{!}{
 \begin{tabular}{ccccccccccccc}
  \hline\noalign{\smallskip}
   \multirow{2}{*}{\begin{tabular}[c]{@{}c@{}}Predicted time \\ horizons
   \end{tabular}} & 
   \multicolumn{2}{c}{LSTM-GPR} & 
   \multicolumn{2}{c}{Persistence} \\ & RMSE (arcsec) & $R^{2}$ & RMSE(arcsec) & $R^{2}$ \\ 
  \hline\noalign{\smallskip}
20 mins & 0.12 & 0.86 & 0.19 & 0.62 \\ 25 mins & 0.13 & 0.74 & 0.20 & 0.49 \\ 30 mins & 0.14 & 0.72 & 0.21 & 0.46 \\ 35 mins & 0.15 & 0.66 & 0.22 & 0.41 \\ 40 mins & 0.18 & 0.54 & 0.23 & 0.39 \\ 45 mins & 0.19 & 0.52 & 0.24 & 0.36 \\ 50 mins & 0.21 & 0.42 & 0.25 & 0.31 \\ 55 mins & 0.23 & 0.36 & 0.25 & 0.29 \\ 60 mins & 0.24 & 0.30 & 0.26 & 0.27 \\ 
\noalign{\smallskip}\hline
\end{tabular}
}

\end{table}

\subsection{Running time considerations}
\label{Running time comparison}


Our proposed LSTM-GPR was implemented in MATLAB 2020a and conducted using an NVIDIA GeForce RTX2080 GPU on a Windows Server with Intel Xeon E5-
2683 v4@2.10GHz CPU and 64GB of RAM. The training time was 7.5 minutes, and the prediction time for one sample was 0.3 seconds. \cite{Yanez-Morroni18} used the WRF model to retrieve daily operational forecasts at a resolution of 6 km, taking 2.5 hours on a 32-core machine. WRF provides numerous meteorological parameters at the vertical level and does well in long-term seeing prediction, while our proposed method has a fairly short calculation time and superior accuracy in short-term seeing prediction. Therefore, the forecaster can use these two methods as complementary strategies for her/his prediction tasks \citep{lyman2020forecasting}.

\section{DISCUSSION AND CONCLUSION}
\label{sec:Dis}

Although numerical models predict that large values of atmospheric seeing ($\gtrsim$1.5 arcsec) within the boundary layer are common over Antarctica's ice sheet \citep{Swain06}, excellent free atmospheric seeing of 0.3 arcsec has been observed at the local elevation maxima on the Antarctic plateau, such as Dome A, C, and F \citep{Ma20a,Lawrence04,Okita13}. The existence of a strong temperature inversion and weak katabatic winds \citep{Hu14} cause a very thin (10--30 m) but stable boundary layer at these local elevation maxima. As a result, most of the optical turbulence that contributes to the integration of $C_n^2(h)$ is confined within this thin boundary layer. Therefore, it is possible to estimate the seeing from the Domes of Antarctica using a multi-layer AWS with several sensors deployed at different heights over a few tens of meters. 

Furthermore, one can forecast the seeing by predicting the meteorological parameters within the boundary layer. Although the mesoscale NWP models might be able to capture thin boundary layers by adjusting the spatial resolution, this would lead to a long calculation time of the model and consume a lot of computational resources, which cannot meet the demand of real-time telescope scheduling.

In this paper, we proposed a novel machine learning-based framework that combines the LSTM neural network with the GPR to estimate and predict seeing at the height of 8 m at Dome A, Antarctic, using only multi-layer meteorological data. LSTM is used to predict future meteorological data, while GPR is used to calculate seeing from predicted meteorological parameters. It is found that the seeing at Dome A is strongly correlated with four meteorological parameters, and the RMSE of the seeing estimation is only 0.18 arcsec. This provides a new way to estimate the seeing at Dome A, without requiring operating a DIMM on-site. The RMSE of seeing prediction 20 minutes into the future is good, at 0.12 arcsec for the seeing range from 0 to 2.2 arcsec. Moreover, since the seeing forecast can be done in less than 1 second, this method is suitable for the real-time scheduling of a telescope.

We also find that the accuracy of the seeing prediction decreases when the temporal fluctuation of the seeing becomes large (the zoomed-in figure in Fig. \ref{fig: seeing}).   
The possible reason for this phenomenon is that the weather data used in this paper only comes from one weather station. In reality the seeing is influenced by the spatial distribution of weather around the site. The accuracy of our proposed model could be further improved if the spatial weather information can be predicted in advance and used as input.
Therefore, we recommend installing gridded weather stations around the observatory. This method has already been confirmed at Cerro Paranal and Cerro Armazones \citep{Milli20}, and
shows that the prediction of seeing peaks is significantly improved.

In this study, only one month's data is used for seeing estimation and seeing prediction. This might impact the algorithm's ability to predict seeing for different times of the year. In the future, we plan to carry out long-term seeing and meteorological measurements simultaneously and to improve our model by training it on night-time data. In addition, since the seeing is also related to spatial meteorological data, we use different data sources to explore the generality and optimization of the algorithm.

\section*{Data Availability}
The seeing and meteorological data at Dome A in 2019 are available in the China-VO Paper Data Repository, at http://paperdata.china-vo.org/BinMa/DomeA-seeing2019.zip. 
The data from KLAWS-2G in 2015 are available at http://aag.bao.ac.cn/klaws/downloads/.

\section*{Acknowledgements}
The authors thank all members of the 27th, 28th, 31st, 32nd, 33rd, and 35th Chinese Antarctic Research Expedition teams for their effort in setting up the KLAWS, KLAWS-2G, and KL-DIMM instruments and servicing the PLATO-A observatory.
This work has been supported by the National Natural Science Foundation of China (NSFC) under grant No. U1831111 and the Operation, Maintenance, and Upgrading Fund for Astronomical Telescopes and Facility Instruments, budgeted from the Ministry of Finance of China (MOF) and administrated by the Chinese Academy of Sciences (CAS). PLATO-A has been supported by Astronomy Australia Limited and the Australian Antarctic Division.
We acknowledge support from the NSFC under grant Nos. 11873010, 11733007, and 12133010, and the Nebula Talents Program of the National Astronomical Observatories, Chinese Academy of Sciences.


\bibliographystyle{elsarticle-harv}
\bibliography{example}


%
%
%

\end{document}